\documentclass[10pt, a4paper]{iopart} 

\usepackage{iopams}
\usepackage{cite}
\expandafter\let\csname equation*\endcsname\relax
\expandafter\let\csname endequation*\endcsname\relax
\usepackage{amsthm}

\usepackage{booktabs}
\usepackage[utf8]{inputenc}
\usepackage[T1]{fontenc}
\usepackage{blindtext}
\usepackage{amsfonts}
\usepackage{amsmath}
\usepackage{enumitem}
\usepackage{microtype}
\usepackage[usenames,dvipsnames]{xcolor}
\definecolor{shadecolor}{gray}{0.9}
\usepackage{graphicx}
\usepackage{overpic}
\usepackage{booktabs}
\usepackage{tabularx}
\usepackage{lipsum}
\usepackage{lengthconvert}
\usepackage{layouts}
\usepackage[sort&compress,numbers]{natbib}
\usepackage[colorlinks,linktoc=all]{hyperref}
\usepackage[all]{hypcap}
\usepackage{breqn}
\definecolor{darkblue}{rgb}{0.0, 0.0, 0.55}
\definecolor{darkmidnightblue}{rgb}{0.0, 0.2, 0.4}
\definecolor{dukeblue}{rgb}{0.0, 0.0, 0.61}
\definecolor{zaffre}{rgb}{0.0, 0.08, 0.66}
\hypersetup{citecolor=blue}
\hypersetup{linkcolor=blue}
\hypersetup{urlcolor=blue}
\usepackage[nameinlink,capitalize]{cleveref}
\usepackage[nameinlink,capitalize]{cleveref}
\Crefname{figure}{Figure}{Figures}
\crefname{figure}{Figure}{Figures}
\Crefname{equation}{Equation}{Equations}

\usepackage{pgf}

\usepackage{caption}
\usepackage{subcaption}
\usepackage{amsthm}
\usepackage{algorithm}
\usepackage{algpseudocode}
\usepackage{scalerel}

\newcommand{\bea}{\begin{eqnarray}}
\newcommand{\eea}{\end{eqnarray}}

\newcommand{\f}[2]{\frac{#1}{#2}}

\newcommand{\ccup}[1]{\left\{#1\right\}}

\newcommand{\Prob}{\mathbb{P}}

\newcommand{\KL}[2]{\mathrm{KL} \left( #1 \, || \, #2 \right)}

\DeclareMathOperator*{\argmax}{arg\,max}

\newcommand{\mess}[3]{q_{#1 \rightarrow #2}(#3)}
\newcommand{\messtwo}[3]{\widehat{q}_{#1 \rightarrow #2}(#3)}
\newcommand{\interquant}{\psi} 
\newcommand{\interprime}{\eta}

\newcommand{\suchthat}{:}
\newcommand{\loglik}{\mathcal{L}}
\newcommand{\pjoint}{p_{H}}
\newcommand{\phye}{p_{E}}
\newcommand{\ppair}{p_{\mathcal{C}}}

\newtheorem{theorem}{Theorem}
\newtheorem{proposition}{Proposition}
\newtheorem*{theorem*}{Theorem}
\newtheorem{lemma}{Lemma}

\newtheorem*{definition*}{Definition}

\usepackage{adjustbox}
\definecolor{shadecolor}{gray}{0.9}
\usepackage{tabu}

\usepackage[normalem]{ulem} 
\usepackage{soul}

\newcommand{\const}{\mathrm{const.}}

\newcommand{\algoname}{HySBM}

\newcommand{\avgdeg}{{d_0}}

\begin{document}

\title[Ruggeri, Lonardi, De Bacco]{Message-Passing on Hypergraphs: Detectability, Phase Transitions and Higher-Order Information}

\author{Nicol{\`o} Ruggeri$^{1,2,*}$, Alessandro Lonardi$^{1,*}$ and Caterina De Bacco$^1$}
\address{$^1$Max Planck Institute for Intelligent Systems, Cyber Valley, T{\"u}bingen 72076, Germany}
\address{$^2$Department of Computer Science{$\mathrm{,}$} ETH{$\mathrm{,}$} Z{\"u}rich 8004{$\mathrm{,}$} Switzerland}
\address{$^*$Equal contribution}
\eads{\mailto{nicolo.ruggeri@tuebingen.mpg.de}, \mailto{\\alessandro.lonardi@tuebingen.mpg.de},\mailto{\\caterina.debacco@tuebingen.mpg.de}}

\date{\today}

\begin{abstract}
Hypergraphs are widely adopted tools to examine systems with higher-order interactions.  Despite recent advancements in methods for community detection in these systems, we still lack a theoretical analysis of their detectability limits.  Here, we derive closed-form bounds for community detection in hypergraphs. Using a Message-Passing formulation, we demonstrate that detectability depends on hypergraphs' structural properties, such as the distribution of hyperedge sizes or their assortativity. Our formulation enables a characterization of the entropy of a hypergraph in relation to that of its clique expansion, showing that community detection is enhanced when hyperedges highly overlap on pairs of nodes.  We develop an efficient Message-Passing algorithm to learn communities and model parameters on large systems. Additionally, we devise an exact sampling routine to generate synthetic data from our probabilistic model.  With these methods, we numerically investigate the boundaries of community detection in synthetic datasets, and extract communities from real systems.  Our results extend the understanding of the limits of community detection in hypergraphs and introduce flexible mathematical tools to study systems with higher-order interactions.
\end{abstract}

\maketitle

\section{Introduction}
\label{sec:intro}

Modeling complex systems as graphs has broadened our understanding of the macroscopic features that emerge from the interaction of individual units. Among the various aspects of this problem, community detection stands out as a fundamental task, as it provides a coarse-grained description of a network's structural organization. Notably,  community structure is observed across different systems, such as food webs \cite{girvan2002community}, spatial migration and gene flow of animal species \cite{fletcher2013network}, as well as in social networks \cite{newman2001structure}, power grids \cite{shekhtman2015resilience},  and others \cite{fortunato2010community}.

In the case of networks with only pairwise interactions, there are solid theoretical results on detectability limits, describing whether the task of community detection can or cannot succeed \cite{decelle2011asymptotic,decelle2011inference,moore2017computer, abbe2018community,ghasemian2016detectability,taylor2016enhanced}. 
However, many complex systems with interactions that extend beyond pairs are better modeled by hypergraphs \cite{battiston2020networks}, which generalize the simpler case of dyadic graphs. Phenomena that have been investigated on graphs are now readily explored on hypergraphs, with examples including diffusion processes, synchronization, phase transitions \cite{battiston2021physics} and, more recently, community structure  \cite{zhou2006learning,chodrow2021generative,contisciani2022principled,ruggeri2023community,dumitriu2023exact}.

Extending the rigorous results of detectability transitions for networks to higher-order interactions is a relevant open question.

One of the main obstacles in modeling hypergraphs is their intrinsic complexity, which poses both theoretical and computational challenges and restricts the range of results available in the literature.
The difficulty of defining communities in hypergraphs and of deriving theoretical thresholds for their recovery has limited investigations to the study of $d$-uniform hypergraphs, i.e., hypergraphs that only contain interactions among exactly $d$ nodes \cite{angelini2015spectral,chien2018community,liang2021information,pal2021community,zhang2022exact,gu2023weak,cole2020exact,chung2017fundamental,ghoshdastidar2017consistency}.

A related line of literature focuses on the detection of planted sub-hypergraphs \cite{yuan2021information,corinzia2022statistical} and testing for the presence of community structure in hypergraphs \cite{jin2021sharp,yuan2021testing}. Generally, extracting recovery results on non-uniform hypergraphs proved to be demanding, with scarce literature on the subject. 

	Recently, Chodrow \emph{et al.} \cite{chodrow2022nonbacktracking} conjectured a recoverability threshold for their spectral clustering algorithm on non-uniform hypergraphs. Closer to the scope of our work, Dumitriu and Wang \cite{dumitriu2023exact} provide a probabilistic model and bounds for the theoretical recovery of communities under the same model.  However, such detectability bounds are based on algorithms which are not feasible in practice, and no empirical demonstration of the predicted recovery is provided.   Furthermore, all these methods lack a variety of desirable probabilistic features,  such as the estimation of marginal probabilities of a node to belong to a community, a principled procedure to sample synthetic hypergraphs with prescribed community structure, and the possibility to investigate the energy landscape of a problem via free energy estimations.

In this work, we address these issues by deriving a precise detectability threshold for hypergraphs that depends on the node degree distribution, the assortativity of the hyperedges, and crucially, on higher-order properties such as the distribution of hyperedge sizes. Additionally, we show how these properties can be formally described via notions of entropy and information, leading to a clear interpretation of the role of higher-order interaction in detectability. 

Our approach is based on a probabilistic generative model and a related Bayesian inference procedure, which we utilize to study the limits of the community detection problem using a Message-Passing (MP) formulation \cite{pearl1982reverend, mezard2009information, murphy2012machine}, originating from the cavity method in statistical physics \cite{mezard1986Spin, mezard2001bethe}.
We focus on an extension to hypergraphs of the stochastic block model (SBM) \cite{holland1983stochastic,wasserman1994social}, a generative model for networks with community structure. Several variants of the SBM \cite{chodrow2021generative}, and of its mixed-membership version \cite{contisciani2022principled,ruggeri2023community}, have been extended to hypergraphs. The model we utilize is an extension of the dyadic SBM to hypergraphs and allows generalizing the seminal detectability results of Decelle \emph{et al.} \cite{decelle2011asymptotic, decelle2011inference} to higher-order interactions.

In addition to our theoretical contributions, we derive an algorithmic implementation for  inferring both communities and parameters of the models from the data.  Our implementation scales well to both large hypergraphs and large hyperedges, owing to a dynamic-program formulation. 

Finally, we show how, with additional combinatorial arguments, one can efficiently sample hypergraphs with arbitrary communities from our probabilistic model. This problem, often studied in conjunction with inference, deserves its own attention when dealing with hypergraphs, as recently discussed in related work \cite{kaminski2022hypergraph,ruggeri2023framework}. 

Through numerical experiments, we confirm our theoretical calculations by
showing that our algorithm accurately recovers the true
community structure in synthetic hypergraphs all the way down
to the predicted detectability threshold.  We also illustrate that our approach gives insights into the community organization of real hypegraphs by analyzing a dataset of group interactions between students in a school.
To facilitate reproducibility, we release open source the code that implements our inference and sampling procedures \cite{GITHUBREPO}.

\section{The hypergraph stochastic block model}
\label{sec: model description}

Consider a hypergraph $H = (V, E)$ where $V = \{1,...,N\}$ is the set of nodes and $E$ the set of hyperedges. A hyperedge $e$ is a set of two or more nodes. We define $\Omega = \{e \suchthat 2 \le |e| \le D\}$, the set of all possible hyperedges up to some maximum dimension $D \le N$, with $|e|$ being the size of a hyperedge, i.e., the number of nodes it contains. Notice that $E \subseteq \Omega$. We denote with $A_e = 1$ all $e \in E$ and with $A_e = 0$ hyperedges $e \in \Omega \setminus E$.

Our Hypergraph Stochastic Block Model (\algoname{}) is an extension of the classical SBM for graphs \cite{holland1983stochastic,wasserman1994social}. It partitions nodes into $K$ communities by assigning a hard membership $t_i \in [K] \equiv \{1, \ldots, K\}$ to each node $ i \in V$,  with $t = \{ t_i \}_{i \in V}$ being the membership vector. It does so probabilistically, assuming that the likelihood to observe a hyperedge $A_{e}$ is a Bernoulli distribution with a parameter that depends on the memberships $\ccup{t_{i}}_{i\in e}$ of its nodes.Formally, the probabilistic model is summarized as
\begin{alignat}{2}
\label{eq: prior}
t_i &\sim \mathrm{Cat}(n) &&\forall i \in V \\
\label{eq: likelihood}
A_e \, | \, t  &\sim \mathrm{Be}\left(\frac{\pi_e}{\kappa_{|e|}}\right) \quad&&\forall e \in \Omega \, ,
\end{alignat}
where $n=(n_1, \ldots, n_K)$ is a vector of prior categorical probabilities for the hard assignments $t_i$. The Bernoulli probabilities are given by
\begin{equation}
\label{eq: pi formula}
\pi_e = \sum_{i < j \in e} p_{t_i t_j} \, ,
\end{equation}
with $0 \leq p_{ab} \leq 1$ being elements of a symmetric probability matrix (also referred to as affinity matrix) and $\kappa_{|e|}$ a normalizing constant that only depends on the hyperedge size $|e|$. This can take on any values, provided that it yields sparse hypergraphs where $\pi_e / \kappa_{|e|} = O(1/N)$ and valid probabilities $\pi_e / \kappa_{|e|}$. We develop our theory for a general form of $\kappa_{|e|}$ and elaborate more on its choice in \ref{apxsec:model}. In our experiments we utilize the value $\kappa_d = \binom{N-2}{d-2} \frac{d(d-1)}{2}$ \cite{ruggeri2023framework, ruggeri2023community}.

Our specific formulation of the likelihood is only one among many alternatives to model communities in hypergraphs. The likelihood we propose has three main properties. First, \algoname{} reduces to the standard SBM when only pairs are present (as $\kappa_2 = 1$). Since we aim to develop a model that generalizes the SBM to hypergraphs, this is an important condition to satisfy.  Second, it enables to develop the MP equations presented in the following section, which in turn lead to a theoretical characterization of the detectability limits and a computationally efficient algorithmic implementation.  Third,  the likelihoods based on expressions similar to \cref{eq: pi formula} have been shown to well describe higher-order interactions that possibly contain nodes from different communities \cite{ruggeri2023framework}. 

For convenience, we work with a rescaled affinity matrix $c = N p$, which is of order $c=O(1)$ in sparse hypergraphs. The log-likelihood $\loglik \equiv \loglik(A, t \,| \, p, n)$ evaluates to
\begin{align}
\loglik &= \sum_{e \in \Omega} \left[A_e \log\left(\frac{\pi_e}{\kappa_e}\right) + (1-A_e) \log\left(1- \frac{\pi_e}{\kappa_e}\right)\right] \nonumber + \sum_{i \in V} \log n_{t_i} \nonumber \\
\begin{split}
 &= \sum_{e \in \Omega} \Bigg[ A_e \log\Bigg(\sum_{i < j \in e} c_{t_i t_j}\Bigg) + (1- A_e) \log\left(1- \frac{\sum_{i < j \in e} c_{t_i t_j}}{N \kappa_e}\right) \Bigg] + \sum_{i \in V} \log n_{t_i} + \const \,, \label{eq: loglik}
 \end{split}
\end{align}
where $\const$ denotes quantities that do not depend on the parameters of the model.

\section{Inference and generative modeling}
\label{sec: inference and generative modeling}

\subsection{Induced factor graph representation}
\label{subsec: factor graph representation}

\begin{figure}[t]
\centering
\includegraphics[width=0.7\columnwidth]{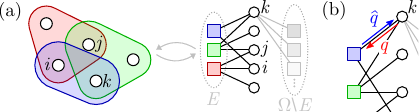}
\caption{
    \textbf{Representing hypergraphs as factor graphs.}
    \textbf{(a)} We depict a hypergraph and its factor graph equivalent. In the factor graph $\mathcal{F}$, function nodes represent hyperedges. Notice that, while the node sets are the same in both representations, due to the presence of all possible hyperedges in the log-likelihood in \Cref{eq: loglik}, the factor graph does not only contain the observed interactions $E$ (black), but also the unobserved ones $\Omega \setminus E$ (gray). 
    \textbf{(b)} In factor graphs, there are two types of messages: variable-to-function node $q$ (red), and function-to-variable node $\hat q$ (blue).
}
\label{fig: hypergraph as factor graph}
\end{figure}

The probabilistic model in \crefrange{eq: prior}{eq: likelihood} has a negative log-likelihood that can be interpreted as the Hamiltonian of a Gibbs-Boltzmann distribution on the community assignments $t$:
\begin{equation}
    \label{eq: gibbs_distribution}
    p(t \, | \, A, p, n) = \frac{p(A, t \, | \, p, n)}{p(A \, | \, p, n)} = \frac{\exp{\loglik(A, t \, | \, p, n)}}{Z} \, ,
\end{equation}
where $Z$ is the partition function of the system, that corresponds to the marginal likelihood of the data. The quantity $F = -\log Z$ is also called the free energy.
The equivalence in \Cref{eq: gibbs_distribution} allows interpreting the probabilistic model in terms of factor graphs \cite{mezard2009information}. Here, the function nodes are hyperedges $f \in \Omega$, and variable nodes are elements of $V$. The interactions between function and variable nodes can be read directly from the log-likelihood in \cref{eq: loglik}. 
In other words, the probabilistic model induces a factor graph $F = (\mathcal{V}, \mathcal{F}, \mathcal{E})$ with variable nodes $\mathcal{V} = V$, function nodes $\mathcal{F} = \Omega$ and edges $\mathcal{E} = \{ (i,e) \in \mathcal{V} \times \mathcal{F} \suchthat i \in e\}$.
 In \cref{fig: hypergraph as factor graph} we show a graphical representation of the equivalence between hypergraphs and factor graphs.
For any variable node $i$ and function node $f$ of the factor graph we define the neighbors, or boundaries, as
$\partial i =\{ f \in \mathcal{F} \suchthat  (i,e) \in \mathcal{E} \}$, being all function nodes adjacent to $i$, and $\partial f =\{ i \in \mathcal{V} \suchthat  (i,e) \in \mathcal{E} \}$ being all variable nodes adjacent to $f$.

\subsection{Message-Passing (MP)}
\label{subsec: message passing}
Given the factor graph representation of \algoname{}, we can perform Bayesian inference of the community assignments via message-passing.
Originally obtained from the cavity method on spin glasses \cite{mezard1986Spin, mezard2001bethe}, MP allows estimating marginal distributions on the variable nodes of a graphical model by iteratively updating messages, auxiliary variables that operate on the edges of the factor graph.  The efficiency of MP comes from the fact that the structure of the factor graph favors locally distributed updates. Although exact theoretical results are only proven on trees, MP has been shown to obtain strong performance also on locally tree-like graphs \cite{mezard2009information} and it has been extended to dense graphs with short loops \cite{cantwell2019message,kirkley2021belief}.

Applying MP to our model, the inference procedure yields expressions for the marginal probabilities $q_i(a)$ of a node $i$ to be assigned to any given community $a \in [K]$. Their values are obtained as solutions to closed-form fixed-point equations, which involve messages $\mess{i}{e}{t_i}$ from variable to function nodes, and $\messtwo{e}{i}{t_i}$, from function to variable nodes. The messages follow the sum-product updates
\begin{align}
\label{eq:MP_eq1}
\mess{i}{e}{t_i} &\propto n_{t_i}  \prod_{f \in \partial i \setminus e} \messtwo{f}{i}{t_i} \\
\label{eq:MP_eq2}
\messtwo{e}{i}{t_i} &\propto \sum_{t_j: j \in \partial e \setminus i} \left( \frac{\pi_e}{\kappa_e} \right)^{A_e} \left( 1- \frac{\pi_e}{\kappa_e} \right)^{1-A_e} \prod_{j \in \partial e \setminus i} \mess{j}{e}{t_j} \, ,
\end{align}
and yield marginal distributions as
\begin{equation}
    \label{eq:MP_marginal}
    q_{i}(t_i) \propto n_{t_i} \prod_{e \in \partial i} \messtwo{e}{i}{t_i} \, .
\end{equation}
Notice that, compared to those for graphs, the MP equations for hypergraphs in \Crefrange{eq:MP_eq1}{eq:MP_marginal}
present additional challenges. First,  in graphs the updates simplify. One can in fact collapse the two types of messages (and equations) into a unique one, since paths $(i, f, j)$ in the factor graph reduce to pairwise interactions $(i, j)$ between nodes. This simplification is not possible in hypergraphs, as one function node may connect more than two variable nodes. Second, the dimensionality of the MP equations grows faster when accounting for higher-order interactions. Here, the number of function nodes is equal to $|\mathcal{F}| = |\Omega| = \sum_{d=2}^D \binom{N}{d}$, yielding $|\mathcal{F}| = O(2^N)$ at large $D=N$. In contrast, one gets $O(N^2)$ pairwise messages in the updates for graphs. To produce computationally feasible MP updates one can assume sparsity, as already done in the dyadic case. We outline such updates in the following theorem. 

\begin{theorem}
\label{th: MP equations}
Assuming sparse hypergraphs where $c=O(1)$, the MP updates satisfy the following fixed-point equations to leading order in $N$. For all hyperedges $e \in E$ and nodes $i \in e$, the messages and marginals are given by:
\begin{align}
    \label{eq:feasible_MP_1}
    {q}_{i \to e}(t_i) &\propto n_{t_i} \Bigg( \prod_{\substack{f \in E \\ f \in \partial i \setminus e}} \widehat{q}_{f \to i}{(t_i)} \Bigg) \exp(-h(t_i)) \\
    \label{eq:feasible_MP_2}
    \widehat{q}_{e \to i}(t_i) &\propto \sum_{t_j: j \in \partial e \setminus i} \pi_e \prod_{j \in \partial e \setminus i} q_{j \to e}{(t_j)} \\
    \label{eq:feasible_MP_3}
    q_i(t_i) &\propto n_{t_i} \Bigg( \prod_{\substack{f \in E \\ f \in \partial i}} \widehat{q}_{f \to i}{(t_i)}  \Bigg) \exp(-h(t_i)) \\
    \label{eq:feasible_MP_4}
    h(t_i) &= \frac{C'}{N} \sum_{j \in V} \sum_{t_j} c_{t_i t_j} q_j(t_j)  \, ,
\end{align}
where $C' = \sum_{d=2}^D \binom{N-2}{d-2} \frac{1}{\kappa_d}$. 
\end{theorem}
A proof of \cref{th: MP equations} is provided in \ref{apxsec:message passing derivations}. The updates in \Crefrange{eq:feasible_MP_1}{eq:feasible_MP_4} are in principle computationally feasible, as products of function nodes $f \in E$ have replaced products over the entire space $f \in \Omega$. In sparse graphs, that we observe in many real datasets, $E$ is much smaller than the original $\Omega$, thus significantly decreasing the computation cost. An intuitive justification of \cref{th: MP equations}, which we formalize in its proof,  is that the observed interactions $f \in E$ hold most of the weight in the updates of their neighbors, while the unobserved ones $f \in \Omega \setminus E$ send approximately constant messages and thus can be absorbed in the external field $h$ introduced in \Cref{eq:feasible_MP_4}. This idea is inspired by the dyadic MP equations in Decelle \emph{et al.} \cite{decelle2011asymptotic}. However, in contrast to MP on graphs, a vanilla implementation of the updates is still not scalable in hypergraphs, as the computational cost of \Cref{eq:feasible_MP_2} is $O(K^{|e|-1})$. To tackle this issue, we develop a dynamic programming approach that reduces the complexity to $O(K^2|e|)$. Dynamic programming is exact, as it does not rely on further approximations on the MP updates,  its detailed derivations are provided in \ref{apxsec: dynamic programming}.

The fixed-point equations of \cref{th: MP equations} naturally suggest an algorithmic implementation of the MP inference procedure. We present a pseudocode for it in \ref{apxsec:implementation details}.

\subsection{Expectation-Maximization to learn the model parameters}
\label{subsec: expectation maximization}

We have presented a MP routine for inferring the community assignments $\ccup{t_{i}}_{i\in V}$. Now, we derive closed-form updates for the model parameters $c,n$ via an Expectation-Maximization (EM) routine \cite{dempster1977maximum}. Differentiating the log-likelihood in \cref{eq: loglik} with respect to $n$, and imposing the constraint $\sum_{a=1}^K n_a = 1$, yields the update
\begin{equation}
\label{eq: n update}
    n_a = \frac{N_a}{N} \, .
\end{equation}
Notice that this update depends on the MP results, as $N_a = |\{i \in V \suchthat \argmax_b q_i(b) = a \}|$ is the count of nodes assigned to community $a$ according to the inferred marginals. To update the rescaled affinity $c$ we adopt a variational approach, where we maximize a lower bound of the log-likelihood, or, equivalently, minimize a variational free energy. In \ref{apxsec: em derivations}, we show detailed derivations for the following fixed-point updates
\begin{equation}\label{eq: c update}
    c_{ab}^{(t+1)} = c_{ab}^{(t)} \, \frac{
        2 \, \sum_{e \in E} {\#^{e}_{ab}} / {\pi_e}
    }{
        N \, C' \, (N n_a n_b - \delta_{ab} n_a)
    } \, ,
\end{equation}
where $\#^{e}_{ab} = \sum_{i < j \in e} \delta_{t_i a} \delta_{t_j b}$ is the count of dyadic interactions between two communities $a, b$ within a hyperedge $e$.
In practice, when inferring $t, n, c$ one proceeds by alternating MP inference of $t$, as presented in \cref{subsec: message passing}, with the updates of $c$ and $n$ in \crefrange{eq: n update}{eq: c update} until convergence. A pseudocode for the EM procedure is presented in \ref{apxsec:implementation details}.

\subsection{Sampling from the generative model}

\label{subsec: sampling}
One of the main advantages of using a probabilistic formulation is the ability to generate data with a desired community structure. Among other tasks, this can be used in particular to test detectability results like the ones we theoretically derive in the following section. However, in hypergraphs, writing a probabilistic model does not directly imply the ability to sample from it, as is typically the case for graphs \cite{ruggeri2023framework, kaminski2022hypergraph}.
In fact, while the $O(N^2)$ configuration space of graphs allows performing sampling explicitly, in the case of hypergraphs the exploding configuration space $\Omega$ makes this task prohibitive, even for hypergraphs with moderate number of nodes and hyperedge sizes. 

We propose a sampling algorithm that can efficiently scale and produce hypergraphs of dimensions in the tens or hundreds of thousands of nodes. We exploit the hard-membership nature of the assignments to obtain exact sampling via combinatorial arguments, as opposed to the approximate sampling in recent work for mixed-membership models \cite{ruggeri2023framework}. The key observation to obtain an efficient algorithm is that the hyperedge probabilities do not depend on the nodes they contain, but only on their community assignments, as implied by \Cref{eq: pi formula}.

With this in mind, we define the auxiliary quantity
\begin{equation}
    \#^e_a = \sum_{i \in e} \delta_{t_i a} \, ,
\end{equation}
for a hyperedge $e$ and community $a \in [K]$, which is the count of nodes in $e$ that belong to community $a$. Crucially, the hyperedge probability depends only on these counts:
\begin{equation}\label{eq: pi in function of counts}
    \pi_e = \sum_{a < b \in [K]} \#^e_a \, \#^e_b \, p_{ab}  + \sum_{a \in [K]} \frac{\#^e_a (\#^e_a - 1)}{2} \, p_{aa} \, .
\end{equation}
Therefore, all hyperedges with different nodes, but same counts $\#^e_1, \ldots, \#^e_K$, have equal probability.

Using \Cref{eq: pi in function of counts}, we sample hypergraphs as in \Cref{pseudocode: Sampling} with the following steps:
\begin{enumerate}
    \item Iterate over the combinations.\\
For hyperedges of size $d=2$, sample all the $N(N-1)/2$ edges directly. Otherwise, iterate the steps (ii), (iii), (iv) for the hyperedge sizes $d = 3, \ldots, D$ and vectors $\# = (\#_1, \ldots, \#_K)$ of community counts (where we omitted the superscript $e$ to highlight that same counts yield identical \Cref{eq: pi in function of counts}) satisfying $\sum_{a = 1}^K \#_a = d$. 
    \item Compute the probability.\\
For a given count vector $\#$, the hyperedge probability $\pi_{\#}$ is given in \cref{eq: pi in function of counts}. Notice that there are $N_{\#} = \binom{N_1}{\#_1} \cdot \ldots \cdot \binom{N_K}{\#_K}$ hyperedges satisfying the count $\#$, since we can choose $\#_a$ nodes from the $N_a$ nodes in each community $a$.
\item Sample the number of hyperedges.\\
Importantly, we do not sample the individual hyperedges, but the \textit{number} of observed hyperedges. Since the individual hyperedges are independent Bernoulli variables with same probability, their sum $X$ follows a binomial distribution:
\begin{equation}\label{eq: binomial count of samples hyperdges}
    X \sim \text{Binom}\left(
        N_{\#}, \frac{\pi_{\#}}{\kappa_d}
    \right) \, 
\end{equation}
with probability $\pi_{\#}$ fixed, determined by $\#$, and number of realizations $N_{\#}$. Sampling directly from \cref{eq: binomial count of samples hyperdges} is numerically challenging for large $N_{\#}$ and $\kappa_d$, hence we adopt a series of numerical approximation summarized in \ref{apxssec: comp coplexity}.
\item Sample the hyperedges.\\
Given the count $X$ of hyperedges sampled from \Cref{eq: binomial count of samples hyperdges}, we can sample the hyperedges. This operation is performed by independently sampling $X$ times $\#_a$ nodes from each community $a$. Notice that this procedure might yield repeated hyperedges, which are not allowed. In sparse regimes, this event has low probability \cite{chodrow2020configuration}. As a sensible approximation, we delete repeated hyperedges.
\end{enumerate}

Owing to this sampling procedure, our results are not limited to theoretical derivations, but can be tested numerically on synthetic data, as we show in \ref{apxssec: experiments sampling}. In \ref{apxssec: comp coplexity} we give a detailed analysis of the complexity, which is asymptotically upper bounded by $O(N \log N)$. A pseudocode for this procedure is shown in \Cref{pseudocode: Sampling} and we provide an open source implementation of the sampling procedure~\cite{GITHUBREPO}.

\begin{algorithm}[htpb]
\caption{\centering Sampling hypergraphs}\label{pseudocode: Sampling}
\begin{algorithmic}
\State {\bf Inputs:} $D$, maximum size of hyperedges
\State \hspace{13mm} $N$, number of nodes
\State \hspace{13mm} $K$, number of communities
\State \hspace{13mm} $n$, prior of the community memberships
\State \hspace{13mm} $p$, affinity matrix
\State 
\State sample node memberships using \cref{eq: prior}
\For{$d=2, \ldots, D$} \Comment{(i)}
    \If{$d=2$}
        \State sample $N(N-1)/2$ (hyper)edges from \cref{eq: likelihood}
    \Else
    	\For{each $\# = (\#_{1},\dots,\#_{K})$ such that $\sum_{a=1}^K \#_a = d$} \Comment{(i)}
    	\State {compute $\pi_\#$ with \Cref{eq: pi in function of counts}} \Comment{(ii)}
    	\State {sample $X$ from \Cref{eq: binomial count of samples hyperdges}} \Comment{(iii)} 
	  \For{$a=1, \ldots, K$}
    	\State {sample $X$ times $\#_a$ nodes} \Comment{(iv)}
    	\EndFor
    	\EndFor
    \State delete repeated hyperedges
    \EndIf
\EndFor
\end{algorithmic}
\end{algorithm}

\section{Phase transition}
\subsection{Detectability bounds}
\label{subsec: detectabilty bounds}

Beside providing a valid and efficient inference algorithm, one of the main advantages of MP is the possibility of deriving closed-form expressions for the detectability of planted communities. The transition from detectable to undetectable regimes has been first shown to exist in MP-based inference models for graphs \cite{decelle2011asymptotic}, and gave rise to an extensive body of literature on theoretical detectability limits and sharp phase transitions \cite{moore2017computer, abbe2018community}. Here, we extend these classical arguments to hypergraphs, and find relevant differences when higher-order interactions are considered.

In line with previous work, we restrict our study to the case where groups have constant expected degrees. In fact, in settings where such an assumption does not hold, it is possible to obtain good classification by simply clustering nodes based on their degrees \cite{decelle2011asymptotic}. Formally, we assume
\begin{equation}\label{eq: constant degree assumption}
    \sum_{b = 1}^K c_{ab}n_b = c \, ,
\end{equation}
for some fixed constant $c$. Notice that \cref{eq: constant degree assumption} does not immediately imply a constant degree for the groups, as in hypergraphs the expected degree is defined differently than the left-hand-side of the equation above. Nevertheless, in \ref{apxsec:proof of constant degree and fixed points} we prove that imposing the condition in \cref{eq: constant degree assumption} does indeed imply a constant average degree. More precisely,
\begin{proposition}
\label{th: constant degree and message fixed points}
    Assuming \cref{eq: constant degree assumption}, the following holds:
    \begin{itemize}
        \item all the groups have the same expected degree;
        \item the fixed points for the messages read
        \begin{alignat}{2}
            \label{eq: fixed_point_1}
		  \mess{i}{e}{t_i} &= n_{t_i} \quad&& \forall e \in E, i \in e \\
            \label{eq: fixed_point_2}
		  \messtwo{e}{i}{t_i} &= \frac{1}{K}  && \forall e \in E, i \in e \, .
	   \end{alignat}
    \end{itemize}
\end{proposition}

\begin{figure}[t]
    \centering
    \includegraphics[width=1\textwidth]{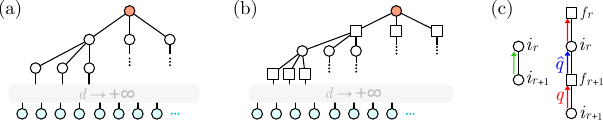}
    \caption{
    \textbf{Local tree assumption.}
    \textbf{(a)} The classical local tree assumption for graphs. Here, it is assumed that the neighborhoods of nodes are approximately trees.
    \textbf{(b)} The tree assumption for factor graphs. Here, a path from a leaf (light blue) to a root (orange) consists of steps alternating variable nodes and function nodes. These two representations coincide in the case of graphs.
    \textbf{(c)} The perturbations propagate up the tree via the messages. In graphs (a), they reach the root passing from nodes $i_{r+1}$ to $i_{r}$ (green). In hypergraph-induced factor graphs, perturbations spread from a node $i_{r+1}$, at depth $r+1$, to its neighboring function nodes $f_{r+1}$ (red), and up to node $i_r$ at depth $r$ (blue) in an alternating fashion.
    }
    \label{fig:factor graph tree assumption}
\end{figure}

We want to study the propagation of perturbations around the fixed points of \Crefrange{eq: fixed_point_1}{eq: fixed_point_2}.
We assume that the factor graph is locally tree-like, i.e., neighborhoods of nodes are approximately trees. We provide a visualization of this in  \cref{fig:factor graph tree assumption}. Classically, it has been proven that for sparse graphs almost all nodes have local tree-like structures up to distances of order $O(\log N)$ \cite{mezard2009information}. We are not aware of similar statements for hypergraphs. While our empirical results prove that these assumptions are reasonable and approximately valid, we leave the formalization of such an argument for future work. 

Referring to \Cref{fig:factor graph tree assumption}(b), one can see that between every leaf and the root, there is a single connecting path. Thus, perturbations on the leaves propagate through a tree to the root, and transmit via the following transition matrix
\begin{equation}
\tilde T^{ab}_r = \frac{\partial \mess{i_r}{f_r}{a}}{\partial \mess{i_{r+1}}{f_{r+1}}{b}} \, ,
\end{equation}
where $i_r, f_r$ are respectively the $r$-th variable node and function node in the path. In words, this is the dependency of a message on the message one level below in the path. In \ref{apxsec: phase transition derivations} we show that, to leading terms in $N$, the transition matrix evaluates to 
\begin{equation}
\label{eq: transition matrix}
    \tilde T^{ab}_r = \frac{2\, n_a}{|f_r|(|f_r|-1)} \left( \frac{c_{ab}}{c} -1 \right) \, .
\end{equation}
A related expression was previously obtained for the transition matrix on graphs is $T^{ab} = n_a \left( {c_{ab}}/{c} -1 \right)$ \cite{decelle2011asymptotic}. Hence, we can compactly write $\tilde T^{ab}_i = [2 / {(|f_r|(|f_r|-1))}] \,T^{ab}$. This connection highlights an important difference between the two cases: hyperedges induce a higher-order prefactor with a ``dispersion'' effect. The larger the hyperedge, the lower is the magnitude of this transition. Instead, if the hyperedge is a pair, this prefactor reduces to one, and we recover the result on graphs. A perturbation $\epsilon_{t_d}^{k_d}$ of a leaf node $k_d$ influences the perturbation $\epsilon_{t_0}^{k_0}$ on the root ${t_0}$ by
\begin{equation}
    \epsilon_{t_0}^{k_0} = \sum_{\{t_r\}_{r=1, \ldots, d}} \left(\prod_{r=0}^{d-1}\tilde T^{t_rt_{r+1}}_i\right) \epsilon_{t_d}^{k_d} \,.
\end{equation}
We can also express this connection in matrix form as
\begin{equation}
    \label{eqn:perturbation_step}
    \epsilon^{k_0}= \left( \prod_{r=0}^{d-1} \frac{2}{|f_r|(|f_r|-1)} \right) T^d \epsilon^{k_d} \, ,
\end{equation}
where $T$ is the matrix with entries $T^{ab}$ (in \Cref{eqn:perturbation_step} raised to the power of $d$), and $\epsilon^{k_d}$ the array of $\epsilon^{k_d}_{t_d}$ values.
Now, similarly to Decelle \emph{et al.} \cite{decelle2011asymptotic}, we consider paths of length $d \rightarrow +\infty$. In such a case, the $r$-dependent prefactor in \Cref{eqn:perturbation_step} converges almost surely to
\begin{align}
\mu = \exp \left( \mathbb{E} \left[d \log \frac{2}{|f|(|f|-1)} \right]  \right) \,,
\end{align}
where the expectation is taken with respect to randomly drawn hyperedges $f \in E$. If $\lambda$ is the leading eigenvector of $T$, then
\begin{equation}
\epsilon^{k_0} \approx \mu \, \lambda^d \epsilon^{k_d} \, .
\end{equation}
Aggregating over the leaves, and since the perturbations have an expected value of zero, we obtain variance:

\begin{align}
    \langle (\epsilon_{t_0}^{k_0})^2 \rangle 
        &\approx \left \langle \left(
            \sum_{k=1}^{[\avgdeg (F-1)]^d} \mu \, \lambda^d  \epsilon^{k}_t
        \right)^2 \right \rangle \\
        \label{eq:last_expression_before_stability}
        &\overset{\text{i.i.d.}}{=} ( \avgdeg (F-1))^d \mu^2\, \lambda^{2d} \langle (\epsilon_{t}^{k})^2  \rangle  \, ,
\end{align}

where $\avgdeg$ is the average node degree and $F$ the average hyperedge size. The expression in  \Cref{eq:last_expression_before_stability} yields the following stability criterion, the key result of our derivations:

\begin{equation}
\label{eq: stability criterion}
\avgdeg (F-1) \left( \exp  \mathbb{E} \left[\log \frac{2}{|f|(|f|-1)} \right]  \right)^2 \lambda^{2} < 1 \, .
\end{equation}

This generalizes the seminal result $c \lambda^2 < 1$ of Decelle \emph{et al.} \cite{decelle2011asymptotic} to hypergraphs. When \Cref{eq: stability criterion} holds, the influence of the leaves to the root decays when propagating up the tree in \Cref{fig:factor graph tree assumption}(b). Conversely, if \Cref{eq: stability criterion} is not satisfied, it grows exponentially.

To obtain more interpretable bounds, we focus on a benchmark scenario where the affinity matrix contains all equal on- and off-diagonal elements, i.e., $c_{aa} = c_{\mathrm{in}}$ for all $a \in [K]$ and $c_{ab} = c_{\mathrm{out}}$ for all $a \neq b$. In this case, condition \cref{eq: constant degree assumption} becomes $c_{\mathrm{in}} + (K-1)c_{\mathrm{out}} = K c$, the leading eigenvalue of $T$ is $\lambda = (c_{\mathrm{in}} - c_{\mathrm{out}}) / Kc$, and the stability condition in \cref{eq: stability criterion} reads

\begin{equation}
\label{eq: bound cin - cout}
    |c_{\mathrm{in}} - c_{\mathrm{out}}| > {\frac{Kc}{\sqrt{\avgdeg (F-1)}}} \exp  \left( - \mathbb{E} \left[\log \frac{2}{|f|(|f|-1)} \right]  \right) \, .
\end{equation}

When hypergraphs only contain dyadic interactions, \Cref{eq: bound cin - cout} reduces to the bound $|c_{\mathrm{in}} - c_{\mathrm{out}}| > K \sqrt{c}$ previously derived for graphs \cite{decelle2011asymptotic}, also known as Kesten-Stigum bound \cite{kesten1967limit, kesten1966additional}.

\subsection{Phase transition in hypergraphs}
\label{subsec: phase transition in hyg}

\begin{figure}[t]
    \centering
    \includegraphics[width=0.55\columnwidth]{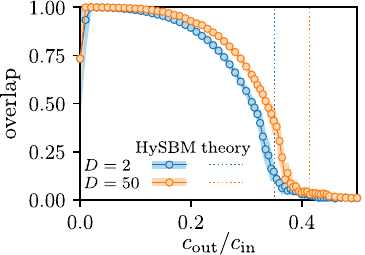}
    \caption{
    \textbf{Phase transition.} The overlap between ground truth and inferred communities varies for different $c_{\mathrm{out}}/c_{\mathrm{in}}$ ratios. The values attained are positive on the detectable region (left of the dotted theoretical bounds) and continuously drop to zero as the phase transition boundary approaches. Values for hyperedges up to size $D=50$ (orange) always yield higher overlap compared to $D=2$ (light blue). Shaded areas are standard deviations over $5$ random initializations of MP.
    }
    \label{fig:phase_transition}
\end{figure}

We test the bound obtained in \Cref{eq: bound cin - cout} by running MP on synthetic hypergraphs generated via the sampling algorithm of \Cref{subsec: sampling}. In our experiments, we fix $K = 4$ and sample hypergraphs with $N = 10^4$ nodes. We also fix $c=10$ and change the ratio $c_{\mathrm{out}}/c_{\mathrm{in}}$. In this setup, for graphs, one expects a continuous phase transition between two regimes where the system is undetectable and detectable \cite{decelle2011asymptotic}. In the former, where the inequality yielded by the Kesten-Stigum bound does not hold, and the graph does not carry sufficient information about the community assignments, community detection is impossible. In the latter, communities can be efficiently recovered by MP. In \Cref{fig:phase_transition} we plot the $\mathrm{overlap} = (\sum_i q_i^\star / N - \max_a n_a) / (1 - \max_a n_a)$ with $q_i^\star \equiv q_i(a_i^\star)$ and $a_i^\star = \argmax_b q_i(b)$, against $c_{\mathrm{out}}/c_{\mathrm{in}}$. Our results are in agreement with the theoretical predictions: the overlap is low in the undetectable region, high in the detectable region, and we observe a continuous phase transition at the Kesten-Stigum bound for graphs, i.e., when $D=2$.

We expect the presence of higher-order interactions to improve detectability, as it yields greater overlap for any $c_{\mathrm{out}}/c_{\mathrm{in}}$ and it shifts the theoretical transition to larger values. We empirically validate this prediction by evaluating \Cref{eq: bound cin - cout} for hyperedges up to size $D=50$ and performing MP inference in \Cref{fig:phase_transition}. Diverging convergence times for larger $c_{\mathrm{out}}/c_{\mathrm{in}}$, i.e., when the free energy landscape gets progressively rugged, further demonstrate this behavior, as shown in \ref{apxsec: Elapsed time of MP}.

\subsection{The impact of higher-order interactions on detectability}
\label{subsec: entropy and higher order info}

As mentioned above, the transition matrix in \cref{eq: transition matrix} reduces to the classic $T^{ab}$ \cite{decelle2011asymptotic} when only dyadic interactions are present. In fact, the additional prefactor ${2} / {(|f_r|(|f_r|-1)})$ is equal to one for 2-dimensional hyperedges.
However, when hyperedges of higher sizes are present, this prefactor is strictly smaller than one. This dampens the perturbations $\epsilon^{k_0}$ when they propagate up the tree in \cref{fig:factor graph tree assumption}(b). It is unclear whether this higher-order effect aids or hurts detectability, as it could prevent signal from being propagated, but also noise from accumulating at the root.

With this in mind, we investigate the impact of higher-oder interactions on detectability by disentangling the effect that $K$, $c$ and, most importantly, $D$ have on the detectability bound set by \Cref{eq: stability criterion}. To this end, we rewrite \Cref{eq: bound cin - cout} as
\begin{equation}
\label{eq:bound using effective deg}
    \left| \rho_{\mathrm{in}} - \frac{1}{Kc} \right| > \Phi(K,c,D) \,. 
\end{equation}
Here, we utilized $c_{\mathrm{in}} / Kc \ = \rho_{\mathrm{in}} \in [0, 1]$, a degree-independent rescaling of $c_{\mathrm{in}}$, where we normalize by its maximum possible value $Kc$, as per \Cref{eq: constant degree assumption}. The term $\Phi(K,c,D)$ is the value of the theoretical bound at the r.h.s. of \Cref{eq: bound cin - cout}, normalized by $Kc$ as well. This way, we get the decomposition $\Phi(K,c,D) = \alpha(K) \beta(c) \gamma(D)$ as a product of three independent terms:
\begin{align}
    \alpha(K) &= \frac{K-1}{K} \label{eq: bound factor 1} \\
    \beta(c) &= \frac{1}{\sqrt{c}} \label{eq: bound factor 2} \\
    \label{eq: gamma_term_detectability}
    \gamma(D) &= \frac{\exp  \left( - \mathbb{E} \left[\log \frac{2}{|f|(|f|-1)} \right]  \right)}{\sqrt{C (F - 1) / 2}} \, ,
\end{align}
where $C = \sum_{d=2}^D \binom{N-2}{d-2} \frac{d}{\kappa_d}$ 

In our experiments we choose of $\kappa_d = \binom{N-2}{d-2} \frac{d(d-1)}{2}$, which conveniently returns $C= 2 H_{D-1}$ (see \ref{apxsec:model}), with $H_{D-1}$ being the $(D-1)$-th harmonic number. However, our theory holds true for any $\kappa_d$ yielding sparse hypergraphs.

\begin{figure}[t]
    \centering
    \includegraphics[width=0.75\columnwidth]{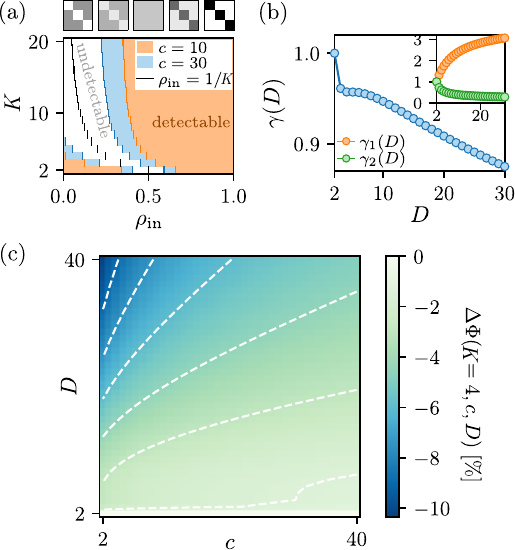}
    \caption{
    \textbf{Theoretical phase transition.}
    Due to the decomposition of our bound in \Crefrange{eq: bound factor 1}{eq: gamma_term_detectability} it is possible to separately describe the effects of $K$, $c$ and $D$ on the predicted phase transition.
    \textbf{(a)} Detectability bounds for networks $(D=2)$. Increasing $c$ yields a broader range of detectable configurations (colored areas) for $\rho_{\mathrm{in}}$. The number of communities skews detectability: while for $K=2$ communities can be detected in extremely disassortative regimes ($\rho_{\mathrm{in}}$ close to zero), when more communities are present, only assortative networks are detectable.
    \textbf{(b)} Effect of the maximum hyperedge size $D$. The term $\gamma(D)$ in \Cref{eq: gamma_term_detectability} can be split into the product $\gamma_1(D)\gamma_2(D)$, as defined in \Crefrange{eq: gamma 1}{eq: gamma 2}. The non-trivial decrease of $\gamma(D)$ results from the interplay of $\gamma_1(D)$ and $\gamma_2(D)$, having opposite monotonicity.
    \textbf{(c)} The percentage decrease $\Delta \Phi(K,c,D) = (\Phi(K,c,D) - \Phi(K,c,2)) / \Phi(K,c,2)$ in detectability for different $c, D$ values shows that higher-order interactions steadily improve detection, especially in sparse regimes.
    }
    \label{fig:detectability_panel}
\end{figure}

The classic effect of $\alpha(K)$ and $\beta(c)$ is summarized in \Cref{fig:detectability_panel}(a), where the maximum hyperedges size is fixed to $D=2$, hence $\gamma(D) = 1$. Here, we observe that the undetectability gap reduces when increasing $c$. Graphs with higher average degrees are more detectable even when there is a larger inter-community mixing. The effect of larger $K$ is that of skewing the detectability phase transition. This is because edges contributing to $c_{\mathrm{out}}$ are spread over $K-1$ communities, while those accounted for $c_{\mathrm{in}}$ concentrate in a single one. Intuitively, increasing $K$ allows to have more in-out edges, and detectability is still possible because of the dominating $c_{\mathrm{in}}$ term. The limit value $\rho_{\mathrm{in}} = 1/K$ constitutes the perfect mixing case $c_{\mathrm{in}} = c_{\mathrm{out}} = c$, where detectability is unfeasible for any $K$ and finite degree $c$. One should notice that, while the bounds drawn in \Cref{fig:detectability_panel} hold theoretically, for large $K$ it may be exponentially hard to retrieve communities even in the detectable region \cite{decelle2011asymptotic,mezard2006reconstruction}.

The higher-order effects on detectability are shown in \Cref{fig:detectability_panel}(b)-(c). The presence of hyperedges with $D > 2$ enters in \Cref{eq: gamma_term_detectability} as the product of two separate contributions, $\gamma(D) = \gamma_1(D) \gamma_2(D)$, where
\begin{align}
    \gamma_1(D) &= \exp  \left( - \mathbb{E} \left[\log \frac{2}{(|f|(|f|-1))} \right] \right) \label{eq: gamma 1} \\
    \gamma_2(D) &= \frac{1}{\sqrt{C(F-1)/2}} \, . \label{eq: gamma 2}
\end{align}
These two terms have contrasting effects that multiply to obtain the overall trend of $\gamma(D)$: $\gamma_1(D)$ is monotonically increasing while $\gamma_2(D)$ is monotonically decreasing. If we were to consider only the ``dispersion'' contribution $\gamma_1$, we would enlarge the detectability gap by increasing $\Phi$. However, the $\gamma_2$ term factors in the increasing number of interactions observed with larger hyperedges.  The result is the overall higher-order contribution to detectability $\gamma(D) = \gamma_1(D) \gamma_2(D)$,  where the value of $\gamma_2$ dominates over $\gamma_{1}$, giving rise to the non-trivial, monotonically decreasing, profile of \Cref{fig:detectability_panel}(b).

The overall effect of higher-order terms is illustrated by plotting the relative difference $\Delta \Phi(K,c,D) = (\Phi(K,c,D) - \Phi(K,c,2)) / \Phi(K,c,2)$ for a range of $c$ and $D$ values, with $K=4$, as shown in \Cref{fig:detectability_panel}(c). We observe how higher-order interactions lead to better detectability for all $c$, especially in sparse regimes, where $c$ is small and pairwise information is not sufficient for the recovery of the communities.

\subsection{Entropy and higher-order information}

Hypergraphs are often compared against their clique decomposition, i.e., the graph obtained by projecting all hyperedges onto their pairwise connections, as a baseline network structure \cite{schneidman2006weak,giusti2015clique,merchan2016sufficiency}.

The clique decomposition yields highly dense graphs. For this reason, most theoretical results on sparse graphs are not directly applicable, algorithmic implementations become heavier---many times unfeasible---and storage in memory is suboptimal. Previous work also showed that algorithms developed for hypergraphs tend to work better in many practical scenarios \cite{contisciani2022principled}. Intuitively, hypergraphs ``are more informative'' than graphs \cite{schneidman2003network}, as there exists only one clique decomposition induced by a given hypergraph, but possibly more hypergraphs corresponding to a given clique decomposition. Here we give a theoretical basis to this common intuition and find that, within our framework, we can quantify the extra information carried by higher-order interactions.

For a given hypergraph $H = (V,E)$, edge $(i, j) \in V^2$ and hyperedge $e \in E$, we define the probability distribution 
\begin{equation}\label{eq:p joint hypergraph}
    \pjoint(\{i, j\}, e) = \begin{cases}
        \displaystyle \frac{1}{E} \frac{2}{|e|(|e|-1)} & \quad \text{if } i, j \in e \\
        0 & \quad \text{otherwise} \,. 
    \end{cases}
\end{equation}
This distribution represents the joint probability of drawing a hyperedge uniformly at random among the possible $E$ in the hypergraph and a dyadic interaction $\{i, j\}$ out of the possible $\binom{|e|}{2}$ within the hyperedge $e$. From \cref{eq:p joint hypergraph} we can derive the following marginal distributions:
\begin{align}
    \phye(e) &= 
        \frac{1}{E} \label{eq: marginal hyperedge distr}\\
    \ppair(\{i, j\}) &= 
        \frac{1}{E}
        \sum_{e \in E: i, j \in e} \frac{2}{|e|(|e|-1)} \, , \label{eq: marginal dyadic distr}
\end{align}
for all $e \in E$ and pairs of nodes $i \neq j$. The distribution $\phye$ is a uniform random draw of hyperedges. The distribution $\ppair$ represents the probability of drawing a weighted interaction $\{i, j\}$ in the clique decomposition of $H$. 

With \Crefrange{eq:p joint hypergraph}{eq: marginal dyadic distr} at hand, it is possible to rewrite $\gamma_1(D)$ in \Cref{eq: gamma 1} as
\begin{align}
    \log \gamma_1(D) 
        &= \mathcal{H}(\{i, j\} \, | \, f) \label{eq: gamma 1 as conditional entropy}\,,
\end{align}

where $\mathcal{H}(\cdot \, | \, \cdot)$ is the conditional entropy. This entropy is minimized when $\ppair(\{i, j\})$ is very different than $ \pjoint(\{i, j\} | f)$, i.e., when conditioning a pair $\{i, j\}$ to be in $f$ brings additional information with respect to the interaction $\{i, j\}$ alone. This happens when $\{i, j\}$ appears in several hyperedges and it is difficult to reconstruct the hypergraph from its clique decomposition. As lower values of $\gamma_1$ imply easier recovery, \Cref{eq: gamma 1 as conditional entropy} suggests that recovery is favored in hypergraphs where  hyperedges overlap substantially and that cannot be easily distinguished from their clique decomposition.

We obtain a similar result by rewriting \Cref{eq: gamma 1 as conditional entropy} as
\begin{equation}\label{eq: gamma 1 as perplexity}
    \gamma_1(D) 
        = \frac{\exp{\mathcal{H}(\pjoint)}}{\exp{\mathcal{H}(\phye)}} 
        = \frac{\mathrm{PP}(\pjoint)}{\mathrm{PP}(\phye)} \,,
\end{equation}	

	which is the ratio of two exponentiated entropies. In information theory, $\mathrm{PP}$ is referred to as perplexity \cite{blei2003latent}, and it is an effective measure of the number of possible outcomes in a probability distribution \cite{campbell1966exponential}. Once we fix the number of hyperedges $E$ (and therefore $\mathrm{PP}(\phye)$),  the number of effective outcomes is given by the number of likely drawn $\{i, j\}$ pairs. This number is minimized when there is high overlap between hyperedges, thus confirming the interpretation of \Cref{eq: gamma 1 as conditional entropy}.

Finally, we set a different focus by rewriting $\gamma_1$ as 
\begin{equation}\label{eq: gamma 1 with KL divergence}
 \log \gamma_1(D) = \mathcal{H}(\ppair) 
            - \KL{\pjoint}{\ppair \, {\otimes} \, \phye} \, ,
\end{equation}
where $\mathrm{KL}$ is the Kullback-Leibler divergence and $\otimes$ the product probability distribution.
Here we pose the question: given a fixed clique decomposition and number of hyperedges, what is the hypergraph attaining the highest detectability? From the equation, such hypergraph is that with the highest $\KL{\pjoint}{\ppair \, {\otimes} \, \phye} = I(\{i, j\}, f)$. In this case, the KL-divergence between a joint distribution and its marginals, also called mutual information $I$ \cite{cover1999elements} of the two random variables, describes the information shared between pairwise interactions and single hyperedges. Hypergraphs with high KL-divergence, i.e, high information about a given $\{i, j\}$ in a single hyperedge $f$, will yield better detectability. In other words, it is preferable to choose hypergraphs that, while still producing the observed clique decomposition (thus achieving low entropy $\mathcal{H}(\pjoint)$), have largely overlapping hyperedges. 
The results discussed in this section provides a theoretically guidance for the construction of hypergraphs that explain an observed graph made of only pairwise interactions  \cite{young2021hypergraph}, a problem relevant in datasets where higher-oder interactions are not explicitly tracked.

\section{Experiments on real data}
\label{sec: experiments real data}
	
\begin{figure}[t]
\centering
\includegraphics[width=1\textwidth]{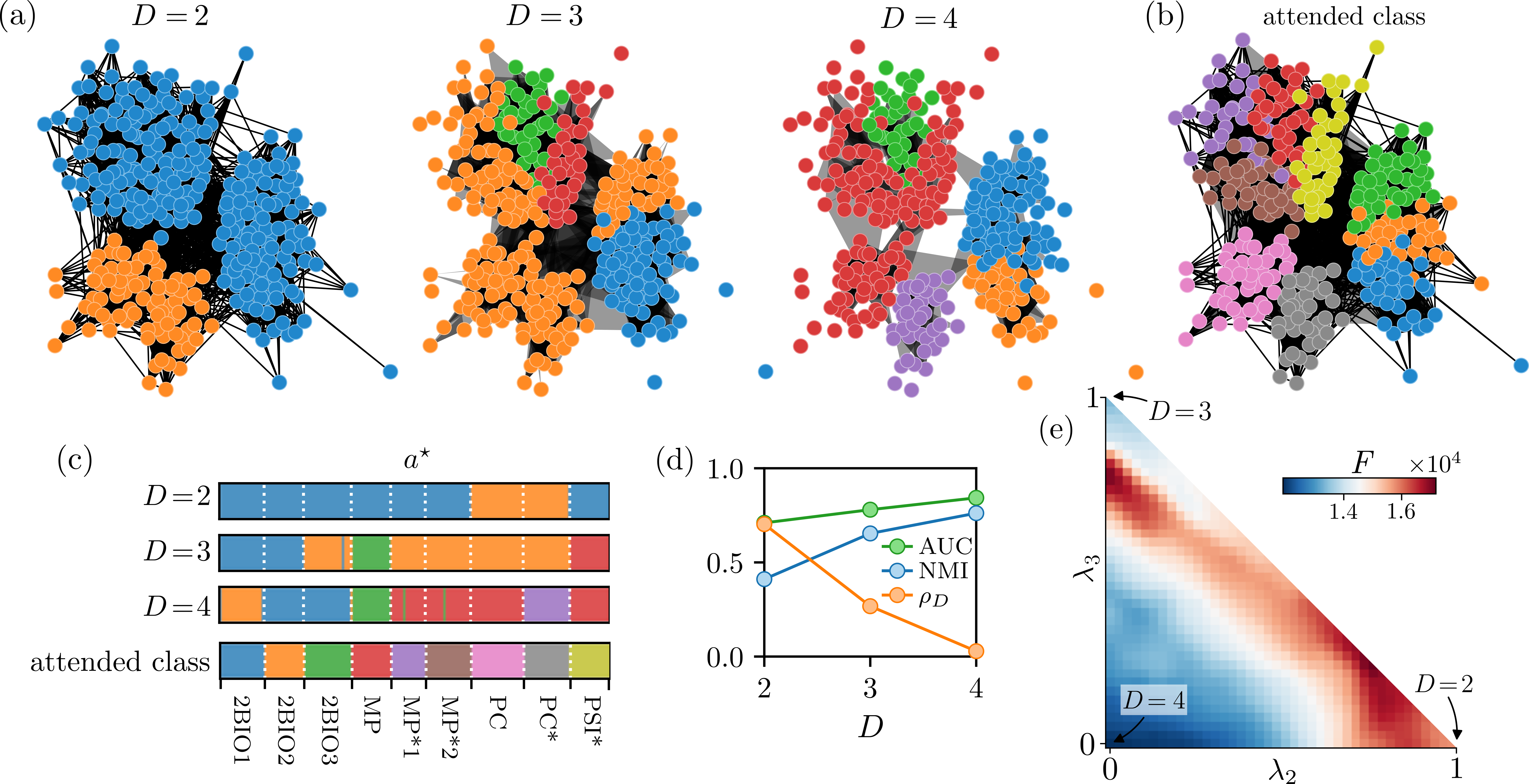}
\caption{\textbf{Experiments on the High School dataset}. We infer the communities via MP and EM on the High School dataset. In all cases, we run inference with $K=10$ communities.
    \textbf{(a)} Inferred communities on the High School dataset, only utilizing hyperedges up to a maximum size $D$. Taking into account higher-order information, up to $D=4$, results in more granular partitions. \textbf{(b)} Graphical representation of the students' partition into classes. We draw only hyperedges of size $D$. \textbf{(c)} We compare the inferred partitions with the ``attended class'' covariate of the nodes, i.e., the classes students participate in.  We comment further on this comparison in \ref{apxsec: affinity on high school}. \textbf{(d)} A quantitative measurement complementing that of panel (b): the Normalized Mutual Information (NMI) between inferred communities and attended classes,  the AUC on the full dataset, as well as the ratio $\rho_D$ of hyperedges of size equal to $D$. \textbf{(e)} Free energy landscape. We consider the parameters $(p_2, n_2)$, $(p_3, n_3)$ and $(p_4, n_4)$ inferred from the dataset with, respectively, $D=2, 3, 4$. With these, we build the simplex of convex combinations $p = \sum_{i \in \{2,3,4\}} \lambda_i  p_i$, where $\sum_{i \in \{2,3,4\}} \lambda_i = 1$ and $0 \leq \lambda_i \leq 1$ (similarly for $n$). For every point in the simplex, we compute the free energy on the full dataset, i.e., with $D=5$. More details on these computations are provided in \ref{apxsec: free energy on high school}.}
\label{fig: real exps}
\end{figure}

Our model leads to a natural algorithmic implementation to learn communities in hypergraphs. In fact,  alternating MP and EM rounds, our algorithm outputs marginal probabilities $q_i(t_i)$ for a node $i$ to belong to a community $t_{i}$, as well as the community ratios $n$ and the affinity matrix $p$. We illustrate an application of this procedure on a dataset of interactions between high school students (High School) \cite{mastrandrea2015contact}. Here, nodes are students and hyperedges represent whether a group of students was observed in close proximity,  as recorded by wearable devices. The hypergraph contains $N=327$ nodes and $E=7818$ hyperedges.
In \Cref{fig: real exps}(a) we show the communities inferred on the dataset where only hyperedges up to size $D=2, 3, 4$ are kept. We observe a clear progression in how the nodes are gradually allocated into different groups when higher-order interactions are progressively taken into account. This suggests that interactions beyond pairs carry information that would get lost if only edges were to be observed. 

To get a qualitative interpretation, we compare the communities inferred with the nine classes attended by the students, an attribute available with the dataset. We illustrate the hypergraph of student interactions, coloring each node according to its class, in \Cref{fig: real exps}(b). Previous studies have shown that in this dataset a number of interactions happen with stronger prevalence within students of the same class \cite{mastrandrea2015contact}. In \Cref{fig: real exps}(c), we compare the communities inferred with different maximum hyperedge size $D$ with the classes, and observe that there is a stronger alignment between them when larger hyperedges are utilized for inference. In \Cref{fig: real exps}(d) we show, at $D=2,3,4$, the Normalized Mutual Information (NMI) between inferred communities and class attributes,the AUC with respect to the full dataset, and the fraction $\rho_D$ of hyperedges with size equal to $D$. In addition, our algorithm detects connection patterns that were previously observed between the different student classes as captured by the affinity matrix $p$, see \ref{apxsec: affinity on high school} for details.

A feature that sets MP apart from other inference methods is the possibility to approximately compute the evidence $Z = p(A \, | \, p, n)$ of the whole dataset, or,  equivalently,  the free energy $F = -\log Z$. In \ref{apxsec: free energy} we discuss how to make the free energy computations feasible by exploiting classical cavity arguments, as well as a dynamic program similar to that employed for MP. We present the results of these estimates on the High School dataset in \Cref{fig: real exps}(e). Here we take the values of $n$ and $p$ inferred by cutting the dataset at maximum hyperedge sizes $D=2, 3, 4$. Then,  we compute the free energy on the full dataset ($D=5$) in the simplex of $n, p$ parameters outlined by the three vertices. We notice that interactions of size $D=5$ seem to be less informative and lead to suboptimal inference, see \ref{apxsec: affinity on high school 2}.
Similarly to what observed on graphs \cite{decelle2011asymptotic}, the energy landscape appears rugged and complex. EM converges to solutions that are local attraction points,  i.e., valleys of low-energy configurations. Moreover, the free energy of the $p, n$ parameters inferred with only pairwise interactions (i.e., $D=2$, lower-right) is higher than that inferred for $D=3$ (upper-left), which is in turn higher that the one of $D=4$ (bottom-left).

\section{Conclusion}
\label{sec: conclusion}
We developed a probabilistic generative model and a message-passing-based inference procedure that lead to several results advancing community detection on hypergraphs. In particular we obtained closed-form bounds for the detectability of community configurations, extending the seminal results of Decelle \emph{et al.} \cite{decelle2011asymptotic} to higher-order interactions. Experimental validation of such bounds shows the emergence of a detectability phase transition when spanning from disassortative to assortative community structures. 
With these theoretical bounds at hand, we investigate the relationship between hypergraphs and graphs from an information-theoretical perspective. Characterizing the entropy and perplexity of pairs of nodes in hyperedges, we find that hypergraphs with many overlapping hyperedges are easier to detect. 
Beside these theoretical advancements, we develop two relevant algorithmic ones. First, we derive an efficient and scalable Message-Massing algorithm to learn communities and model parameters. Second, we propose an exact and efficient sampling routine that generates synthetic data with desired community structure according to our probabilistic model in order of seconds. 
Both of these implementations are released open source \cite{GITHUBREPO}.

The mathematical tools we propose here to obtain our results are valid for standard hypergraphs. We can foresee that they could be generalized to dynamic hypergraphs where interactions change in time, using intuitions derived for dynamic graphs \cite{ghasemian2016detectability}.  Similarly, it would be interesting to see how detectability bounds change when accounting for node attributes, as results in networks have shown that adding extra information can boost community detection \cite{newman2016structure,contisciani2020community,badalyan2023hypergraphs}. Finally, from an empirical perspective, it would be interesting to see how our theoretical insights in terms of entropy of hypergraphs and clique expansion match measures that relate hypergraphs to simplicial complexes \cite{landry2023simpliciality}.

\section*{Acknowledgments}
N.R. and A.L. contributed equally to this work. N.R. acknowledges support from the Max Planck ETH Center for Learning Systems. 
The authors thank the International Max Planck Research School for Intelligent Systems (IMPRS-IS) for supporting A.L.

\appendix



\section{Expected degree and choice of $\kappa_d$}
\label{apxsec:model}
As we commented in \cref{sec: model description}, the choice of the normalizing constant $\kappa_d$, for $d=2, \ldots, D$, controls the Bernoulli probabilities for all hyperedges $e \in \Omega$ via 
\begin{equation}
    \mathbb{P}(e \, | \, p, t) = \frac{\pi_e}{\kappa_{|e|}} = \frac{\sum_{i<j \in e} p_{t_i t_j}}{\kappa_{|e|}} \,.  \nonumber
\end{equation}
Our theoretical analysis and results hold for general choices of $\kappa_d$, as long as these respect the following conditions. First, for any choice of a symmetric $0 \leq p_{ab} \leq 1$, we need valid probabilities $0 \leq \pi_e / \kappa_{|e|} \leq 1$. This implies that, necessarily:
\begin{equation}
\label{eq: minimum kappa}
    \kappa_d \ge \frac{d(d-1)}{2} \quad \forall d=2, \ldots, D \, .  
\end{equation}
Second, we want the ensemble to consist of sparse hypergraphs, in expectation. A good proxy for such a requirement is the average degree, which we can compute explicitly:

\begin{align}
\avgdeg
	&= \frac{1}{N} \sum_{i \in V} \sum_{e \in \Omega: i \in e} \mathbb{P}(e \, | \, p, t)   \nonumber \\
	&= \frac{1}{N} \sum_{e \in \Omega} \sum_{i \in e} \mathbb{P}(e \, | \, p, t)  \nonumber \\
	&= \frac{1}{N} \sum_{e \in \Omega} |e| \mathbb{P}(e \, | \, p, t)  \nonumber \\
	&= \frac{1}{N} \sum_{e \in \Omega} \frac{|e|}{\kappa_{|e|}} \sum_{i< j \in e} p_{t_i t_j}  \nonumber \\
	&= \frac{C}{N} \sum_{i <j \in V} p_{t_i t_j}  \nonumber \\
	&\approx  \frac{C}{N} \sum_{a \le b \in [K]} \frac{p_{ab} (N n_a) (N n_b)}{1 + \delta_{ab}}  \nonumber  \\
	\label{eq: final expression avg deg}
    &= \frac{C}{2} \sum_{a,b \in [K]} c_{ab}n_a n_b \, , 
\end{align}

where 
\begin{equation}\label{eq: C formula}
C = \sum_{d=2}^D \binom{N-2}{d-2} \frac{d}{\kappa_d} \, .  \nonumber
\end{equation}
We assume $c_{ab} = O(1)$, i.e. to be in a sparse regime. Thus, the expected degree's scale is governed by $C$ and, in turn,  by the choice of $\kappa_d$,  as
\begin{equation}
\avgdeg = O(C)  \nonumber \, .
\end{equation}

Additionally, but not necessarily, we wish our model to extend the classical SBM, which imposes the additional condition $\kappa_2 = 1$.
There exist many choices of $\kappa_d$ obeying the constraints just discussed. A natural one is the minimum value satisfying \cref{eq: minimum kappa}, i.e. $\kappa_d = {d(d-1)}/{2}$. This gives
\begin{equation}
    C = \frac{2}{N-1} \sum_{d=1}^{D-1} \binom{N-1}{d} \, \nonumber
\end{equation}
that, for $D=N$, returns $\avgdeg = O \left( {2^{N}} / ({N-1}) \right)$, which is too high to yield sparse hypergraphs. Notice that, in practice, we rarely use $D=N$. However, such considerations are useful to evaluate how different $\kappa_d$ values reflect on the properties of the hypergraph ensembles of the model.

A more interesting choice is given by  
\begin{equation}
\kappa_d = \frac{d(d-1)}{2} \binom{N-2}{d-2} \, .  \nonumber
\end{equation}
This corresponds to taking the average among the ${d(d-1)}/{2}$ interactions that yield $\pi_e$,  and $\binom{N-2}{d-2}$ is a normalization: once observed an interaction between two nodes $i, j$,  the remaining $d-2$ are chosen at random. This gives
\begin{align}
C &= 2 \sum_{d=1}^{D-1} \frac{1}{d} \nonumber \\
\label{eq: kappa choice in appendix}
 &= 2 H_{D-1} \, ,
\end{align}
which is proportional to the $(D-1)$-th harmonic number, hence growing more mildly at leading order as $C = O(\log D)$. Aside from having an interpretation in terms of null modeling, the value in \cref{eq: kappa choice in appendix}, which we utilize experimentally, was shown to be a sensible choice in many real-life scenarios \cite{ruggeri2023framework, ruggeri2023community}.

\section{Message-Passing derivations}
\label{apxsec:message passing derivations}
MP equations have been developed in the case of general factor graphs, see for example Murphy \emph{et al.}~\cite{murphy2012machine}, Section 22.2.3.2.  We consider approximate messages from hyperedges $e$ to nodes $i$ being $\messtwo{e}{i}{t_i}$, and vice versa, $\mess{i}{e}{t_i}$. The messages, for any $e \in \mathcal{F}, i \in \partial e$ satisfy the general updates
\begin{align}
\mess{i}{e}{t_i} &\propto n_{t_i}  \prod_{f \in \partial i \setminus e} \messtwo{f}{i}{t_i} \nonumber \\
\messtwo{e}{i}{t_i} &\propto \sum_{t_j: j \in \partial e \setminus i} \left( \frac{\pi_e}{\kappa_e} \right)^{A_e} \left( 1- \frac{\pi_e}{\kappa_e} \right)^{1-A_e} \prod_{j \in \partial e \setminus i} \mess{j}{e}{t_j} \, . \label{eq: messtwo formula appendix}
\end{align}
The marginal beliefs are given by
\begin{equation}
\label{eq: general marginals appendix}
q_{i}(t_i) \propto n_{t_i} \prod_{e \in \partial i} \messtwo{e}{i}{t_i} \, . 
\end{equation}

\subsection{Message updates}
First, we can distinguish the values of messages for function nodes $e$ such that $A_e=0$ or $A_e=1$, i.e. if the hyperedge $e$ is observed or not in the data.

If $A_e = 1$, i.e. $e \in E$, then 
\begin{align}
\messtwo{e}{i}{t_i}
	&\propto \sum_{t_j: j \in \partial e \setminus i} \frac{\pi_e}{\kappa_e} \prod_{j \in \partial e \setminus i} \mess{j}{e}{t_j} \nonumber\\
	&\propto \sum_{t_j: j \in \partial e \setminus i} \pi_e \prod_{j \in \partial e \setminus i} \mess{j}{e}{t_j} \,.  \label{eq: e to i for present hye}
\end{align}

If $A_e = 0$, then $e \in \Omega \setminus E$. 
We start by computing
\begin{align}
\messtwo{e}{i}{t_i}
	&\propto \sum_{t_j: j \in \partial e \setminus i} \left(
			1 - \frac{\pi_e}{\kappa_e} 		
		\right)
		\prod_{j \in \partial e \setminus i} \mess{j}{e}{t_j} \nonumber \\
	&= \sum_{t_j: j \in \partial e \setminus i} \prod_{j \in \partial e \setminus i} \mess{j}{e}{t_j} 
        - \sum_{t_j: j \in \partial e \setminus i} \frac{\pi_e}{\kappa_e} \prod_{j \in \partial e \setminus i} \mess{j}{e}{t_j} \nonumber \\
	&= 1 - \sum_{t_j: j \in \partial e \setminus i} \frac{\pi_e}{\kappa_e} \prod_{j \in \partial e \setminus i} \mess{j}{e}{t_j} \nonumber \\
	&= 1- \frac{1}{N}\sum_{t_j: j \in \partial e \setminus i} \frac{\sum_{k < m \in e}c_{t_k t_m}}{\kappa_e} \prod_{j \in \partial e \setminus i} \mess{j}{e}{t_j} \, .\label{eq: e to i non existent}
\end{align}
We indicate with $\hat{Z}_{e\to i}(t_i)$ the convenient non-normalized rewriting of $\messtwo{e}{i}{t_i}$ in \Cref{eq: e to i non existent}. Therefore, we find
\begin{align}
\mess{i}{e}{t_i}
	&\propto n_{t_i} \prod_{f \in \partial i \setminus e} \messtwo{f}{i}{t_i} \nonumber \\
     \label{eq: second_step_approximate_message}
    &= \frac{n_{t_i}}{\messtwo{e}{i}{t_i}} \prod_{f \in \partial i} \messtwo{f}{i}{t_i}  \\
    \label{eq: third_step_approximate_message}
	&\propto   \frac{n_{t_i}}{\hat{Z}_{e\to i}(t_i)} \prod_{f \in \partial i} \messtwo{f}{i}{t_i} \\
     &= \frac{q_{i}{(t_i)}}{\hat{Z}_{e\to i}(t_i)} \label{eq: i to e no expansion} \,,
    \end{align}
where from the \Cref{eq: second_step_approximate_message} to \Cref{eq: third_step_approximate_message} we used $\hat{Z}_{e\to i}(t_i)$ introduced in \Cref{eq: e to i non existent}. We evaluate the expression in \Cref{eq: i to e no expansion} for the limit $N \to + \infty$, which gives the node-to-hyperedge messages for $e \in \Omega \setminus E$ as
\begin{align}
    \mess{i}{e}{t_i} &= q_{i}{(t_i)} + O\left( \frac{1}{N} \right) \nonumber \\
    \label{eq: approx i to e}
    &\approx q_{i}{(t_i)} \,,
\end{align}
i.e., the nodes approximately (to leading order in $O(1/N)$) share their marginal belief to hyperedges that are not observed in the data. Using \Cref{eq: approx i to e}, we can also approximate \Cref{eq: e to i non existent} as
\begin{align}
\messtwo{e}{i}{t_i}
	&\propto 1- \frac{1}{N}\sum_{t_j: j \in \partial e \setminus i} \frac{\sum_{k < m \in e}c_{t_k t_m}}{\kappa_e} \prod_{j \in \partial e \setminus i} \mess{j}{e}{t_j} \nonumber \\
	&\approx 1- \frac{1}{N}\sum_{t_j: j \in \partial e \setminus i} \frac{\sum_{k < m \in e}c_{t_k t_m}}{\kappa_e} \prod_{j \in \partial e \setminus i} q_j(t_j) \, .  \label{eq: e to i marginal}
\end{align}
In the assumed sparsity regime,  the term of order $O(1/N)$ in \Cref{eq: e to i marginal} is close to zero.  Since for $x \approx 0$ the approximation $1-x \approx e^{-x}$ is sufficiently accurate, we write
\begin{align}\label{eq: final e to i in exponential}
\messtwo{e}{i}{t_i} &\approx \exp\left(
	-\frac{1}{N}\sum_{t_j: j \in \partial e \setminus i} \frac{\sum_{k < m \in e}c_{t_k t_m}}{\kappa_e} \prod_{j \in \partial e \setminus i} q_j(t_j)
\right) \,.
\end{align}

We can put the hyperedge-to-node updates together using the two results in \Cref{eq: e to i for present hye} and in \Cref{eq: final e to i in exponential}. Specifically, we derive the following expression for the message $q_{i \to e}(t_i)$,  where $e \in E$:
\begin{align}
\mess{i}{e}{t_i}
	&\propto n_{t_i} \prod_{\substack{f \in \Omega: \\ f \in \partial i \setminus e}} \messtwo{f}{i}{t_i} \nonumber \\
	&= n_{t_i} \prod_{\substack{f \in E: \\ f \in \partial i \setminus e}} \messtwo{f}{i}{t_i}  \prod_{\substack{f \in \Omega \setminus E \\ f \in \partial i}} \messtwo{f}{i}{t_i} \nonumber \\
	\label{eq: approx message 2} &\approx n_{t_i} \left( \prod_{\substack{f \in E: \\ f \in \partial i \setminus e}} \widehat{q}_{f \to i}{(t_i)} \right) \left[ \prod_{\substack{f \in \Omega \setminus E: \\ f \in \partial i}} \exp\left(
		-\frac{1}{N}\sum_{t_j: j \in \partial f \setminus i} \frac{\sum_{k < m \in f} c_{t_k t_m}}{\kappa_f} \prod_{j \in \partial f \setminus i} q_j(t_j)
	\right) \right] \\
	&= n_{t_i} \left( \prod_{\substack{f \in E: \\ f \in \partial i \setminus e}} \widehat{q}_{f \to i}{(t_i)} \right) \exp\left(
		-\frac{1}{N} \sum_{\substack{f \in \Omega \setminus E: \\ f \in \partial i}} \sum_{t_j: j \in \partial f \setminus i} \frac{\sum_{k < m \in f} c_{t_k t_m}}{\kappa_f} \prod_{j \in \partial f \setminus i} q_j(t_j)
	\right)  \nonumber \\
 	&\approx n_{t_i} \left( \prod_{\substack{f \in E: \\ f \in \partial i \setminus e}} \widehat{q}_{f \to i}{(t_i)} \right) \exp\left(
		-\frac{1}{N} \sum_{\substack{f \in \Omega: \\ f \in \partial i}} \sum_{t_j: j \in \partial f \setminus i} \frac{\sum_{k < m \in f} c_{t_k t_m}}{\kappa_f} \prod_{j \in \partial f \setminus i} q_j(t_j)
	\right)  
	\label{eq: approx message 4} \\
 &= n_{t_i} \left(  \prod_{\substack{f \in E: \\ f \in \partial i \setminus e}} \widehat{q}_{f \to i}{(t_i)} \right) \exp(-h_i(t_i)) \label{eq: ext field} \,.
\end{align}
In \Cref{eq: approx message 2}, we used the approximation introduced in \Cref{eq: final e to i in exponential}. In \Cref{eq: approx message 4} we passed from summing over $\Omega \setminus E$ to $\Omega$. This approximation is sensible as long as the expected degree of the nodes grows at most as $N$, which is satisfied in the assumed sparse regime, as discussed in \ref{apxsec:model}. Finally, in \Cref{eq: ext field} we introduced node-dependent external field $h_i(t_i)$ whose definition naturally follows from the argument of the exponential in \Cref{eq: approx message 4}.

\subsection{External field updates}
We simplify the external field to remove the node dependency of $h_i(a)$. The node-dependent external field reads
\begin{align}
h_i(t_i) 
	&= \frac{1}{N} \sum_{f \in \partial i} \frac{1}{\kappa_f}
		\left(\sum_{t_j: j \in \partial f \setminus i} \, \,  \sum_{k < m \in f}c_{t_k t_m} \prod_{r \in f \setminus i} q_r(t_r) \right) \nonumber \\
	&= \frac{1}{N} \sum_{f \in \partial i } \frac{1}{\kappa_f}
		\left(\sum_{t_j: j \in f \setminus i} \, \,  \sum_{m \in f \setminus i}c_{t_i t_m} \prod_{r \in f \setminus i} q_r(t_r) \right) + \const \, \label{eq: external field 1}
\end{align}
The sum in parentheses in \Cref{eq: external field 1} can be simplified as
\begin{align}
\sum_{t_j: j \in f \setminus i} \left[ \left( \sum_{m \in f \setminus i}c_{t_i t_m} \right) \prod_{r \in f \setminus i} q_r(t_r) \right] \nonumber &= \sum_{t_j: j \in f \setminus i} \, \,  \sum_{m \in f \setminus i}  \left( c_{t_i t_m} \prod_{r \in f \setminus i} q_r(t_r) \right)   \nonumber \\
	&= \sum_{m \in f \setminus i} \, \,  \sum_{t_j: j \in f \setminus i} \left( c_{t_i t_m} \prod_{r \in f \setminus i} q_r(t_r) \right)   \nonumber \\
	&= \sum_{m \in f \setminus i} \,\,  \sum_{t_m}  c_{t_i t_m} q_m(t_m)  \, . \label{eq: inner summand h_i}	
\end{align}
Plugging \Cref{eq: inner summand h_i} into \Cref{eq: external field 1} we get, ignoring constants,
\begin{align}
h_i(t_i)
	&= \frac{1}{N} \sum_{f \in \partial i} \frac{1}{\kappa_f}
		\sum_{m \in f \setminus i} \,\,  \sum_{t_m}  c_{t_i t_m} q_m(t_m) \nonumber \\
	&= \frac{C'}{N} \sum_{j \in V \setminus  i} \sum_{t_j} c_{t_i t_j} q_j(t_j)  \nonumber \\
	&\approx \frac{C'}{N} \sum_{j \in V} \sum_{t_j} c_{t_i t_j} q_j(t_j) \,, \label{eq: external field supp intermediate}
\end{align}
with $C' = \sum_{d=2}^D \binom{N-2}{d-2} \frac{1}{\kappa_d}$ and where in \Cref{eq: external field supp intermediate} we included $i$ in the node summation. Since \cref{eq: external field supp intermediate} does not depend on $i$, we define the node-independent external field
\begin{equation}
\label{eq: external field final supp}
    h(a) = \frac{C'}{N} \sum_{j \in V} \sum_{t_j} c_{a t_j} q_j(t_j) \quad \forall a \in [K] \,. \nonumber
\end{equation}

\subsection{Marginal beliefs updates}
Notice, that, in passing from \cref{eq: approx message 2} to \cref{eq: ext field} and then in \cref{eq: external field final supp},  we have shown that 
\begin{equation}
    \prod_{\substack{f \in \Omega \setminus E \\ f \in \partial i}} \messtwo{f}{i}{t_i}
        \approx \exp(-h_i(t_i)) \approx \exp(-h(t_i)) \, .
\end{equation}
We use the same argument to treat the general expression of the marginal beliefs in \cref{eq: general marginals appendix}, yielding
\begin{align}
    q_{i}(t_i) 
        &\propto n_{t_i} \prod_{e \in \partial i} \messtwo{e}{i}{t_i} \nonumber \\
        &= n_{t_i} 
            \prod_{\substack{e \in E: \\ e \in \partial i}} \messtwo{e}{i}{t_i} 
            \prod_{\substack{e \in \Omega \setminus E: \\ e \in \partial i}} \messtwo{e}{i}{t_i} \nonumber \\
        &\approx n_{t_i} 
            \prod_{\substack{e \in E: \\ e \in \partial i}} \messtwo{e}{i}{t_i} 
            \exp(-h(t_i)) \, . \nonumber
\end{align}

\subsection{Summary: approximate Message-Passing updates}
Putting all derivations together, the final MP equations read
\begin{alignat}{3}
    \text{Node-to-observed hyperedge:}& \quad \mess{i}{e}{t_i} &&\propto n_{t_i} \left( \prod_{\substack{f \in E \\ f \in \partial i \setminus e}} \widehat{q}_{f \to i}{(t_i)} \right) \exp(-h(t_i)) \quad &&\forall e \in E, i \in e \nonumber \\
    \label{eq: final_mp_costly}
    \text{Observed hyperedge-to-node:}& \quad \messtwo{e}{i}{t_i} &&\propto \sum_{t_j: j \in \partial e \setminus i} \pi_e \prod_{j \in \partial e \setminus i} q_{j \to e}{(t_j)}  \quad &&\forall e \in E, i \in e \\
    \text{External field:}& \quad\quad\;
        h(t_i) &&= \frac{C'}{N} \sum_{j \in V} \sum_{t_j} c_{t_i t_j} q_j(t_j) \label{eq: external field final apx} \\ 
    \text{Marginals:}& \quad\quad\; q_i(t_i) &&\propto n_{t_i} \left( \prod_{\substack{f \in E \\ f \in \partial i}} \widehat{q}_{f \to i}{(t_i)}  \right) \exp(-h(t_i)) \, \nonumber. 
\end{alignat}

Notice that the MP updates cannot be naively implemented as presented. In fact, the update in \Cref{eq: final_mp_costly} for $\messtwo{e}{i}{t_i}$ have cost $O(K^{|e|-1})$, which does not scale with the hyperedge size. In \ref{apxsec: computational details} we present a dynamic programming approach to perform this computation exactly with cost $O(K^2 |e|)$, and comment on further algorithmic details to implement the MP updates in practice.

\section{Expectation-Maximization inference}
\label{apxsec: em derivations}

\textit{Updates of the community priors $n$.} We take the derivative of the log-likelihood in \cref{eq: loglik}. By imposing the constraint $\sum_{a=1}^K n_a = 1$, we obtain the update in \cref{eq: n update}.\\
\textit{Updates of the affinity matrix $p$.} We show here the updates in terms of $c$. These easily translate to those in terms of the affinity matrix $p$ as the expression we derive below in \Cref{eq: em update c} is invariant with respect to the substitution $c = N p$.
 Let $x_e = \sum_{i < j \in e} c_{t_i t_j} /  N \kappa_e$. Then, ignoring additive constants, the log-likelihood reads
\begin{align}
\loglik
	&= \sum_{e \in E} \log \left(
		\sum_{i<j \in e} c_{t_i t_j}
	\right)
	+ \sum_{e \in \Omega \setminus E} \log(1-x_e) \nonumber \\
    \label{eq:approx}
	&\approx \sum_{e \in E} \log \left(
		\sum_{i<j \in e} c_{t_i t_j}
	\right)
	- \sum_{e \in \Omega \setminus E} x_e \\
    &= \sum_{e \in E} \log \left(
		\sum_{i<j \in e} c_{t_i t_j}
	\right)
	- \sum_{e \in \Omega \setminus E} \frac{\sum_{i<j \in e} c_{t_i t_j}}{N \kappa_e} \nonumber 
\end{align}
where \Cref{eq:approx} is the linearization of $\log(1-x) \approx x$ around $x=0$, which is valid at leading order $O(1/N)$. We now take a variational approach to find a lower bound $\tilde{\loglik}$ of the log-likelihood:
\begin{align}
\loglik
	&\approx \sum_{e \in E} \log \left(
		\sum_{i<j \in e} c_{t_i t_j}
	\right)
	- \sum_{e \in \Omega \setminus E} \frac{\sum_{i<j \in e} c_{t_i t_j}}{ N \kappa_e} \nonumber \\
 \label{eq: jensen}
	&\geq \sum_{e \in E} \sum_{i <j \in e}\rho_{ij}^{e} \log \left(
		\frac{c_{t_i t_j}}{\rho_{ij}^{e}}
	\right)
	- \sum_{e \in \Omega \setminus E} \frac{\sum_{i<j \in e} c_{t_i t_j}}{ N \kappa_e} \\
	&= \sum_{e \in E} \sum_{i <j \in e} \rho_{ij}^{e} \log c_{t_i t_j} \nonumber 
	- \sum_{e \in \Omega \setminus E} \frac{\sum_{i<j \in e} c_{t_i t_j}}{ N \kappa_e} + \const\\
	&= \tilde{\loglik}(c) + \const\, , \nonumber 
\end{align}
which is valid for any distribution $\rho_{ij}^{e}$ such that $\sum_{i<j \in e} \rho_{ij}^{e} = 1$. In \Cref{eq: jensen}, we utilized Jensen's inequality. The lower bound is exact when 
\begin{equation}
\label{eq: optimal rho}
    \rho_{ij}^{e} = \frac{c_{t_i t_j}}{\sum_{i< j \in e} c_{t_i t_j}} = \frac{c_{t_i t_j}}{N \pi_e} \quad.
\end{equation}
 
We compute the derivative of the variational lower bound and approximate to leading terms in $N$:
\begin{align}
    \frac{\partial \tilde{\loglik}}{\partial c_{ab}}
        &= \frac{1}{c_{ab}} \sum_{e \in E} \sum_{i < j \in e} \rho_{ij}^{e} \delta_{t_i a} \delta_{t_j b}
            - \frac{1}{N} \sum_{e \in \Omega \setminus E} \frac{1}{\kappa_e} \sum_{i < j \in e} \delta_{t_i a} \delta_{t_j b} \nonumber \\
        &\approx \frac{1}{c_{ab}} \sum_{e \in E} \sum_{i < j \in e} \rho_{ij}^{e} \delta_{t_i a} \delta_{t_j b}
            - \frac{1}{N} \sum_{e \in \Omega} \frac{1}{\kappa_e} \sum_{i < j \in e} \delta_{t_i a} \delta_{t_j b} \label{eq: log-lik derivative indexes} \\
        &= \frac{1}{c_{ab}} \sum_{e \in E} \sum_{i < j \in e} \rho_{ij}^{e} \delta_{t_i a} \delta_{t_j b}
            - \frac{C'}{N} \sum_{i < j \in V} \delta_{t_i a} \delta_{t_j b} \nonumber \\
        &= \frac{1}{c_{ab}} \sum_{e \in E} \sum_{i < j \in e} \rho_{ij}^{e} \delta_{t_i a} \delta_{t_j b}
            - \frac{C'}{2  N} (N_a N_b - \delta_{ab} N_a) \nonumber \\
        &= \frac{1}{c_{ab}} \sum_{e \in E} \sum_{i < j \in e} \rho_{ij}^{e} \delta_{t_i a} \delta_{t_j b}
            - \frac{C'}{2} (N n_a n_b - \delta_{ab} n_a) \, . \label{eq: approx c derivative lower bound}
\end{align}
where $C' = \sum_{d=2}^D \binom{N-2}{d-2} \frac{1}{\kappa_d}$. Notice that the approximations in \Cref{eq: log-lik derivative indexes} and  \Cref{eq: approx c derivative lower bound} hold valid only when considering $c_{ab}$ in the expressions, as by assumption $c = O(1)$. Now, by setting \cref{eq: approx c derivative lower bound} equal to zero, and substituting $\rho_{ij}^{e}$ from \cref{eq: optimal rho}, we obtain the update
\begin{equation}\label{eq: em update c}
    c_{ab}^{(t+1)} = c_{ab}^{(t)} \, \frac{
        2 \, \sum_{e \in E} {\#^{e}_{ab}} / {\pi_e}
    }{
        N \, C' \, (N n_a n_b - \delta_{ab} n_a)
    } \, ,
\end{equation}
where $\#^{e}_{ab} = \sum_{i < j \in e} \delta_{t_i a} \delta_{t_j b}$.

\section{Algorithmic and computational details}
\label{apxsec: computational details}

\subsection{Dynamic programming for MP}
\label{apxsec: dynamic programming}
In this section, we explain how the MP updates for the $\messtwo{e}{i}{t_i}$ messages can be performed efficiently. In log-space, the messages can be compactly written as
\begin{align}
\log \messtwo{e}{i}{t_i} 
	&= \log \sum_{t_j : j \in \partial e \setminus i} \pi_e \prod_{j \in e \setminus i} \mess{j}{e}{t_j} + \const \nonumber \\
	&= \interquant(e, i, t_i) + \const  \label{eq: psi_def}\quad .
\end{align}
Below, we focus on finding efficient updates for $\interquant$ as defined in \Cref{eq: psi_def}, which should be exponentiated and properly normalized to find the original messages $\messtwo{e}{i}{t_i}$.
For this, we introduce an auxiliary quantity. For any subset $g \subseteq f$ of nodes in $f$,  where $i \in g$, we define
\begin{equation}
\interprime(g, i, t_i) = \log \left[
	\sum_{t_j : j \in g \setminus i} 
		\left(\sum_{l < m \in g} p_{t_l t_m} \right) 
		\prod_{j \in g \setminus i} \mess{j}{f}{t_j} 
\right] \, . \nonumber
\end{equation}
Hence $\interprime(f, i, t_i) = \interquant(f, i, t_i) + \const $. This quantity is useful in that it allows to obtain an efficient recursion formula for $\interquant$, by computing the $\interprime$ values starting from subsets $g$ containing two nodes.

Without loss of generality, consider $f=\{1, \ldots, m-1\}$ and $i=1$. Consider $g = \{1, \ldots, n-1\}$ for some $n \le m$. We want to compute $\interprime$ for the set $\{1, \ldots n\}$. Its exponential is given by:
\begin{align}
\fl
\exp\left(
	\interprime(\{1, \ldots, n\}, 1, t_1) 
\right)
	& = \sum_{t_n} \sum_{t_2} \sum_{t_3} \ldots \sum_{t_{n-1}} 
		\left(
			p_{t_1 t_2} + \ldots + p_{t_{n-2} t_{n-1}} + p_{t_1 t_n}  
    +p_{t_2 t_n} + \ldots + p_{t_{n-1} t_n} \right) \nonumber \\ 
    &\times \left(\mess{2}{f}{t_2} \ldots \mess{n-1}{f}{t_{n-1}} \mess{n}{f}{t_n}\right) \nonumber \\
	&= \sum_{t_n} \mess{n}{f}{t_n} \Bigg( \sum_{t_2} \sum_{t_3} \ldots \sum_{t_{n-1}} (p_{t_1 t_2} + \ldots + p_{t_{n-2} t_{n-1}}) (\mess{2}{f}{t_2} \ldots \mess{n-1}{f}{t_{n-1}}) \nonumber \\
		&+  \sum_{t_2} \sum_{t_3} \ldots \sum_{t_{n-1}} p_{t_1 t_n} (\mess{2}{f}{t_2} \ldots \mess{n-1}{f}{t_{n-1}}) \nonumber \\
		&+ \sum_{t_2} \sum_{t_3} \ldots \sum_{t_{n-1}} (p_{t_2 t_n} + \ldots + p_{t_{n-1} t_n}) (\mess{2}{f}{t_2} \ldots \mess{n-1}{f}{t_{n-1}}) \nonumber \Bigg)\nonumber \\
	&= \sum_{t_n} \mess{n}{f}{t_n} \Bigg(  \exp( \interprime(\{1, \ldots, n-1\}, 1, t_1)) \nonumber \\
		&+ p_{t_1 t_n} \nonumber\\
		&+ \sum_{t_2} p_{t_2 t_n}\mess{2}{f}{t_2} + \ldots + \sum_{t_{n-1}} p_{t_{n-1} t_n}\mess{n-1}{f}{t_{n-1}} \Bigg)  \nonumber \\
   \begin{split}
	&= \exp( \interprime(\{1, \ldots, n-1\}, 1, t_1)) \\
		&+ \sum_{t_n} \mess{n}{f}{t_n} \Bigg( p_{t_1 t_n}  + \sum_{t_2} p_{t_2 t_n}\mess{2}{f}{t_2} + \ldots + \sum_{t_{n-1}} p_{t_{n-1} t_n}\mess{n-1}{f}{t_{n-1}} \Bigg)  \, .  \label{eq: recursion formula dynamic}
   \end{split}
\end{align}

The recursion in \cref{eq: recursion formula dynamic} allows to compute the value of $\interprime(\{1, \ldots, n\}, 1, t_1)$ from $\interprime(\{1, \ldots, n-1\}, 1, t_1)$ in time $O((n-2) \, K^2)$. However, we can further reduce the cost. For any $a \in [K]$, define
\begin{equation*}
s_n(a) = \sum_{t_2} p_{t_2 a} \mess{2}{f}{t_2} + \ldots + \sum_{t_{n-1}} p_{t_{n-1} a} \mess{n-1}{f}{t_{n-1}} \, .
\end{equation*}
Substituting the definition of $s_n(a)$ in \cref{eq: recursion formula dynamic}, we obtain the final two-step dynamic update:
\begin{alignat}{2}
&s_n(a) = s_{n-1}(a) + \sum_{t_{n-1}} p_{t_{n-1} a} \mess{n-1}{f}{t_{n-1}} \label{eq: s updates}\\
&\exp\left( \interprime(\{1, \ldots, n\}, 1, t_1) \right) = \exp( \interprime(\{1, \ldots, n-1\}, 1, t_1)) \nonumber \\
	& \qquad\qquad\qquad + \sum_{t_n} \mess{n}{f}{t_n} \left(
		 	p_{t_1 t_n} + s_n(t_n)
		 \right) \, . \label{eq: simplified dynamic formula}
\end{alignat}
This yields a cost of $O(K)$ per recursion, and a total cost of $O(K \, |f|)$ to compute the final $\interquant(f, 1, t_1)$. In practice, for any $e, i$ pair, we compute $\interquant(e, i, t_i)$ for all values $t_i \in [K]$, which yields a total cost of  $O(K^2 \, |f|)$.

\subsection{Implementation details}
\label{apxsec:implementation details}
In our implementation of the MP and EM routines, we take some additional steps to ensure convergence to non-trivial local optima of the free energy landscape.

The initialization of the messages is performed taking into account the circular relationships in \Crefrange{eq:feasible_MP_1}{eq:feasible_MP_4}. We perform them as follows: (i) randomly initialize the messages $\mess{i}{e}{t_i}$. For every $i, e$ pair, the messages are drawn from a $K$-dimensional Dirichlet distribution. (ii) Similarly, randomly initialize the marginal beliefs $q_i(t_i)$. (iii) We infer all the other quantities from the initialized $\mess{i}{e}{t_i}$ and $q_i(t_i)$. In fact, up to constants
    \begin{equation*}
        \messtwo{e}{i}{t_i} = \frac{
            q_i(t_i)
        }{
            \mess{i}{e}{t_i}
        } \, .
    \end{equation*}
All values are then normalized to have unitary sum. 
(iv) Finally, the external field is entirely determined by the marginals, as per \cref{eq:feasible_MP_4}.

We check for convergence of the MP and EM inference routines by evaluating the absolute difference between parameters in consecutive steps. We present complete pseudocodes of the two routines in \cref{pseudocode: MP} and \cref{pseudocode: EM}.

\begin{algorithm}[H]
\caption{\centering Inferring communities (MP)}\label{pseudocode: MP}
\begin{algorithmic}
\State {\bf Inputs:} convergence threshold $\epsilon_{\mathrm{mp}}$
\State \hspace{13mm} maximum iterations $\mathrm{iter}_{\mathrm{mp}}$
\State \hspace{13mm} prior $n$, rescaled affinity matrix $c$
\State 
\State randomly initialize all $\mess{i}{e}{t_i}, \messtwo{e}{i}{t_i}, q_i(t_i), h(t_i)$\\
\For{$\mathrm{step}=1, \ldots, \mathrm{iter}_{\mathrm{mp}}$}
    \State \emph{// Perform updates}
    \For{all $e \in E, i \in e, t_i \in [K]$}
        \State update messages $\mess{i}{e}{t_i}$ \Comment{\cref{eq:feasible_MP_1}}
    \EndFor
    \For{all $e \in E, i \in e, t_i \in [K]$}
        \State update messages $\messtwo{e}{i}{t_i}$ \Comment{\cref{eq:feasible_MP_2}}
    \EndFor
    \For{all $e \in E, i \in e, t_i \in [K]$}
        \State $q_i^{old}(t_i) \gets q_i(t_i)$
        \State update marginals $q_i(t_i)$ \Comment{\cref{eq:feasible_MP_3}}
    \EndFor
    \For{$t_i \in [K]$}
        \State update external field $h(t_i)$ \Comment{\cref{eq:feasible_MP_4}}
    \EndFor
    \State 
    
    \emph{// Check for convergence}
    \State $\Delta = \sum_{i=1}^N \sum_{t_i=1}^K |q_i^{\mathrm{old}}(t_i) - q_i(t_i)|$
    \If{$\Delta < \epsilon_{\mathrm{mp}}$}
        \State break 
    \EndIf
\EndFor
\end{algorithmic}
\end{algorithm}
\begin{algorithm}[H]
\caption{\centering Inferring model parameters (EM)}\label{pseudocode: EM}
\begin{algorithmic}
\State {\bf Inputs:} convergence threshold $\epsilon_{\mathrm{em}}$
\State \hspace{13mm} maximum iterations $\mathrm{iter}_{\mathrm{em}}$
\State 
\State randomly initialize $c, n$\\
\For{$\mathrm{step}=1, \ldots, \mathrm{iter}_{\mathrm{em}}$}
    \State \emph{// Perform updates}
    \State perform Message-Passing inference \Comment{\cref{pseudocode: MP}}
    \State $n^{\mathrm{old}} \gets n$
    \State update $n$ \Comment{\cref{eq: n update}}
    \State $c^{\mathrm{old}} \gets c$
    \State update $c$ \Comment{\cref{eq: c update}}
    \State 
    \State \emph{// Check for convergence}
    \State $\Delta = \sum_{a=1}^K |n_a - n_a^{\mathrm{old}}| + \sum_{a, b=1}^{K} |c_{ab} - c_{ab}^{\mathrm{old}}|$
    \If{$\Delta < \epsilon_{\mathrm{em}}$}
        \State break 
    \EndIf
\EndFor
\end{algorithmic}
\end{algorithm}

While \cref{pseudocode: MP} is presented as a completely parallel implementation of the MP equations \Crefrange{eq:feasible_MP_1}{eq:feasible_MP_4}, in practice we proceed in batches. In fact, we find that applying completely parallel updates, i.e. applying \cref{eq:feasible_MP_1} for all $i, e$ pairs, successively \cref{eq:feasible_MP_2} for all $i, e$ pairs, and then \cref{eq:feasible_MP_3} for all nodes $i \in V$, results in fast convergence to degenerate fixed-points where all nodes are assigned to the same community. For this reason, we apply dropout. Given a fraction $\alpha \in (0, 1]$, we select a random fraction $\alpha$ of all possible $i, e$ pairs, and apply the update in \cref{eq:feasible_MP_1} only for the selected pairs. We perform a new random draw, and update according to \cref{eq:feasible_MP_2}, and similarly for \cref{eq:feasible_MP_3}. Finally, we update the external field in \cref{eq:feasible_MP_4}. 
Empirically, we find that a value of $\alpha=0.25$ works for synthetic data, where inference is simpler. Values below work as well. For real data we find that substantially lowering $\alpha$ yields more stable inference. On real data,  where we alternate MP and EM, and learning is less stable, we utilize $\alpha=0.01$.  In practice, we also set a patience parameter, and only stop MP once a given number of iterations in a row falls below the threshold $\epsilon_{mp}$ in \Cref{pseudocode: MP}. For real datasets, we set the patience to $50$ consecutive steps, and the maximum number of iterations $\text{iter}_{\text{mp}}=2000$. 

\section{Sampling from the generative model}
\label{apxsec: sampling}

\subsection{Computational complexity}
\label{apxssec: comp coplexity}

For a fixed hyperedge size $d$, there are two parts to the computational cost: iterating through the counts $\#$, and sampling the hyperedges. The number of counts is fixed and given by $K^d / d!$, i.e., the number of possible ways to assign $d$ nodes to $K$ groups, without order. This cost corresponds to performing steps (ii) and (iii) of the sampling algorithm in \Cref{subsec: sampling}, where one needs to enumerate all possible counts $\#$,  which are $K^d / d!$ for every dimension $d$, and sample from a binomial distribution for each count. The cost of sampling the hyperedges in step (iv) in \Cref{subsec: sampling} can also be precisely quantified.  Every $d$-dimensional hyperedge is sampled with a computational cost of $d$ since it is exactly the extraction of $d$ nodes from $V$, and there are $\omega_{d}$ of such hyperedges. Calling $\Omega^d$ the space of all $d$-dimensional hyperedges, we find

\begin{align}
    \mathbb{E}[\omega_d] 
	&= \sum_{e \in \Omega^d} \Prob(e \, | \, p , t) \nonumber\\
	&= \sum_{e \in \Omega^d} \sum_{i < j \in e} \frac{p_{t_i t_j}}{\kappa_d} \nonumber\\
	&= \frac{1}{\kappa_d}\sum_{e \in \Omega^d} \sum_{i < j \in e} p_{t_i t_j} \nonumber\\
	&= \frac{\binom{N-2}{d-2}}{\kappa_d} \sum_{i < j \in V} p_{t_i t_j} \nonumber\\
    &\approx \frac{\binom{N-2}{d-2} N^2}{\kappa_d} \sum_{a \le b =1}^K p_{ab} n_a n_b \nonumber\\
    &= \frac{\binom{N-2}{d-2} N}{\kappa_d} \sum_{a \le b =1}^K c_{ab} n_a n_b \, . \nonumber 
\end{align}

Hence, the average computational cost is given by

\begin{equation}
  \label{eq: expected_num_hye_d}
    \sum_{d=2}^D \left(
        \frac{K^d}{d!} +
        d \, \mathbb{E}[\omega_d]
		\right) \, .
\end{equation}

Given the large size of $\Omega^d$, the cost in \Cref{eq: expected_num_hye_d} tightly concentrates around the expected value. In sparse regimes, the term ${K^d}/{d!}$ dominates as the number of hyperedges $\omega_d$ is low, while the two terms both contribute to the cost when $\mathbb{E}[\omega_d]$ grows. 

Precisely, we quantify the cost in \Cref{eq: expected_num_hye_d} in terms of asymptotic complexity.  The first summand $\sum_{d=2}^D \frac{K^d}{d!}$ absolutely converges to a constant for diverging $D$, and contributes to the complexity only as a constant relevant in sparse regimes. Defining $a_{d} = K^{d}/d!$, we can use the ratio test to assess convergence:
\begin{align}
 \lim_{d \to + \infty} \left| \f{a_{d+1}}{a_{d}} \right| = 
\lim_{d \to + \infty} \f{ K^{d+1} d! }{K^d (d+1)d!} = \lim_{d \to + \infty} \f{K}{d+1} = 0 \,.
\end{align}
Substituting the value of $\kappa_d = \frac{d (d-1)}{2} \binom{N-2}{d-2}$ that we utilize in our experiments, it is also possible to quantify the second addend
\begin{align*}
	\sum_{d=2}^D d \, \mathbb{E}[\omega_d] 
		&\approx \sum_{d=2}^D d \left( \frac{\binom{N-2}{d-2} N}{\kappa_d} \sum_{a \le b =1}^K c_{ab} n_a n_b \right) \\
		&= \left( 
				\sum_{a \le b =1}^K c_{ab} n_a n_b 
			\right) \, 
			\left(
				2 N \, \sum_{d={2}}^D \frac{1}{d-1} 
			\right)\, .
\end{align*}
Similar to the reasoning presented in \Cref{eq: kappa choice in appendix}, choosing the maximum possible cost, given by $D=N$ (which is higher than most practical use cases), the sum $\sum_{d={2}}^D \frac{1}{d-1}$ grows like $O(\log N)$, therefore $\sum_{d=2}^D d \, \mathbb{E}[\omega_d] = O(N \, \log N)$, which yields an asymptotic bound of the total sampling complexity.

Finally, we remark that since sampling from \Cref{eq: binomial count of samples hyperdges} is computationally costly, we approximate the binomial with a Gaussian distribution \cite{gauss_corr}, or with a Poisson if $N_{\#}$ is large and ${\pi_{\#}}/{\kappa_d}$ is small \cite{pois_corr}. We use a Ramanujan approximation for large log-factorials appearing in the calculations \cite{raman_approx}.

\subsection{Experiments}
\label{apxssec: experiments sampling}

We employ the sampling algorithm to generate the hypergraphs used to study the phase transition of \cref{subsec: phase transition in hyg}. Here, we set the affinity matrix to have all equal in-degree $c_{aa} = c_{\mathrm{in}}$ and out-degree $c_{ab} = c_{\mathrm{out}}$, so that \cref{eq: constant degree assumption} becomes $c_{\mathrm{in}} + (K-1)c_{\mathrm{out}} = K c$ for some $K$ and $c$. In our experiments, we sample hypergraphs with $N=10^4$ nodes by fixing $c=10$ and $K=4$, we span across $65$ values of $c_{\mathrm{out}}$ in $[0,500]$, and compute the corresponding $c_{\mathrm{in}} = c_{\mathrm{in}}(c_{\mathrm{out}}; K,c)$. For each experimental configuration $c_{\mathrm{in}}, c_{\mathrm{out}}$, we draw $5$ hypergraphs from different random seeds. This gives a total of $325$ hypergraphs.

We use the expected number of $d$-dimensional hyperedges $\mathbb{E}[{\omega}_d]$ in \Cref{eq: expected_num_hye_d} and the average degree $\avgdeg$ in \Cref{eq: final expression avg deg} to perform a sanity check between our sampling algorithm and theoretical derivations. For constant in and out-degree, these two metrics evaluate to

\begin{align*}
    \mathbb{E}[\omega_d] &\approx \frac{N c}{d(d-1)} \quad,\\
    \avgdeg &\approx \frac{C c}{2} \,.
\end{align*}

\begin{figure}[htpb]
\centering
\includegraphics[width=0.5\columnwidth]{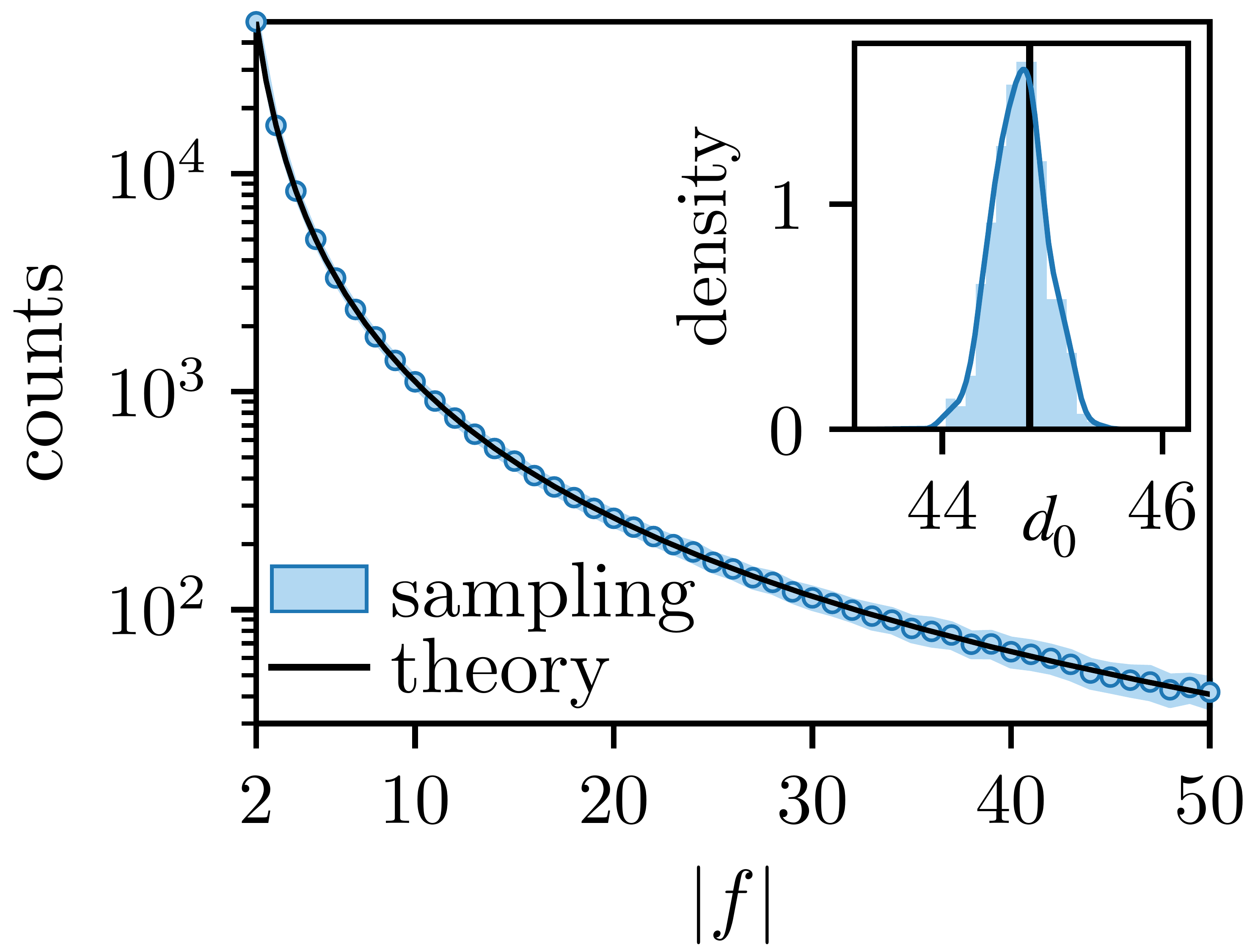}
\caption{\textbf{Sampling experiments.} The expected number of $|f|$-dimensional hyperedges returned by our experiments (blue) is in great accordance with the theoretical prediction $\mathbb{E}[{\omega}_{|f|}]$ (black). Similarly, the experimental expected degrees distribute around the analytical $\avgdeg$. Shaded areas are standard deviations over $5$ random hypergraph extractions, at each $|f|$.}
\label{fig: sampling exps}
\end{figure}

The results in \Cref{fig: sampling exps} show excellent agreement between theory and experiments. We also highlight that the sampling method is extremely fast and has an average sampling time of $t = 32.7 \pm 2.7 \mathrm{(s)}$ on the experimental setup considered here.

\section{Phase transition: complementary derivations and additional results}

\subsection{Proof of \Cref{th: constant degree and message fixed points}}
\label{apxsec:proof of constant degree and fixed points}

First, we want to prove that all communities have the same expected degree. In order to do that, we start by computing the expected degree $ \avgdeg_i $ of a given node $i \in V$. Following similar derivations to those for $\avgdeg$  \ref{apxsec:model}, we find
\begin{align*}
\avgdeg_i
	&= \sum_{e \in E: i \in e} \Prob(e \, | \, p, t) \\
	&= \sum_{e \in E: i \in e} \frac{\pi_e}{\kappa_e} \\
	&= C'  \sum_{a=1}^K c_{t_i a} n_a + 
	\frac{N C''}{2} \left(
		\sum_{b, d=1}^K c_{bd} n_b n_ d
		+ \sum_{b=1}^K c_{bb} n_b^2
	\right) \, ,
\end{align*}
where $C' = \sum_{d=2}^D {\binom{N-2}{d-2}} / {\kappa_d}$, as previously defined, and $C'' = \sum_{d=3}^D {\binom{N-3}{d-3}} / {\kappa_d}$. Therefore, the average degree $\langle b \rangle$ of a community $b \in [K]$, evaluates to

\begin{align*}
\langle b \rangle
	&= \frac{1}{N_b} \sum_{i \in V: t_i = b} \avgdeg_i  \\
	&=  \frac{1}{N_b} \sum_{i \in V: t_i = b} \left[
		C'  \sum_{a=1}^K c_{t_i a} n_a + 
		\frac{N C''}{2} \left(
		\sum_{d, m=1}^K c_{dm} n_d n_ m
		+ \sum_{d=1}^K c_{dd} n_d^2
	\right)
	\right] \\
	&=  \frac{1}{N_b} \sum_{i \in V: t_i = b} \left[
		C'  \sum_{a=1}^K c_{b a} n_a + 
		\frac{N C''}{2} \left(
		\sum_{d,m=1}^K c_{dm} n_d n_ m
		+ \sum_{d=1}^K c_{dd} n_d^2
	\right)
	\right] \\
	&= C'  \sum_{a=1}^K c_{b a} n_a + 
		\frac{N C''}{2} \left(
		\sum_{d,m=1}^K c_{dm} n_d n_ m
		+ \sum_{d=1}^K c_{dd} n_d^2
	\right) \\
	&= C' c  + 
		\frac{N C''}{2} \left(
		\sum_{d,m=1}^K c_{dm} n_d n_ m
		+ \sum_{d=1}^K c_{dd} n_d^2 
	\right)\quad,
\end{align*}

which is independent of the specific choice of group $b$, from which we conclude that all the groups yield equal expected degrees.

Second, we wish to demonstrate that MP's fixed points are as in \Crefrange{eq: fixed_point_1}{eq: fixed_point_2}. Notice that in the derivations here below, when convenient, we interchange equivalent summations over function nodes' neighbors $\partial e$ and hyperedge $e$. By treating all quantities that are independent of $t_i$ in $q_{i \rightarrow e}(t_i), \messtwo{e}{i}{t_i}$ as a constant, we evaluate \Cref{eq:MP_eq2} as
\begin{align}
\messtwo{e}{i}{t_i}
	&\propto \sum_{t_j : j \in e \setminus i} \frac{\pi_e}{\kappa_e} \prod_{j \in \partial e \setminus i} q_{j \rightarrow e}(t_j)  \nonumber  \\
	&\propto \sum_{t_j : j \in e \setminus i} \sum_{r < s \in e} p_{t_r t_s} \prod_{j \in \partial e \setminus i} q_{j \rightarrow e}(t_j)  \nonumber  \\
	&= \sum_{t_j : j \in e \setminus i} 
	\left(
	\sum_{r \in e \setminus i} p_{t_r t_i} \prod_{j \in \partial e \setminus i} q_{j \rightarrow e}(t_j) + 
	\sum_{r<s \in e \setminus i} p_{t_r t_s} \prod_{j \in \partial e \setminus i} q_{j \rightarrow e}(t_j)
	\right) \nonumber   \\
	&= \sum_{r \in e \setminus i} \sum_{t_r} p_{t_r t_i} q_{r \rightarrow e}(t_r)  + 
		\sum_{r<s \in e \setminus i} \sum_{t_r, t_s} p_{t_r t_s} q_{r \rightarrow e}(t_r) q_{s \rightarrow e}(t_s)  \nonumber  \\
	&= \sum_{r \in e \setminus i} \sum_{t_r} p_{t_r t_i} n_{t_r}  + 
		\sum_{r<s \in e \setminus i} \sum_{t_r, t_s} p_{t_r t_s} n_{t_r} n_{t_s}  \nonumber  \\
	&= \frac{1}{N} \left(
            \sum_{r \in e \setminus i} c  + 
		\sum_{r<s \in e \setminus i} c 
        \right)  \nonumber  \\
	&= \frac{c}{N} \left( (|e|-1) + c\frac{|e|(|e|-1)}{2} \right) \, .
    \label{eq: prop_to_constant_mess_e_to_i}
\end{align}
Since messages $\messtwo{e}{i}{t_i}$ are normalized to have unitary sum, \Cref{eq: prop_to_constant_mess_e_to_i} implies that $\messtwo{e}{i}{t_i} = {1}/{K}$. Substituting this result into \Cref{eq:MP_marginal}, one finds also that $q_i(t_i) = n_{t_i}$. The variable-to-function node messages are updated with \Cref{eq:feasible_MP_1}, which includes \Cref{eq:feasible_MP_4} for the external field $h(t_i)$. The external field evaluated at fixed points is also constant, in fact
\begin{align}
h(t_i)
	&= \frac{C'}{N} \sum_{j \in V} \sum_{t_j} c_{t_i t_j} q_j(t_j) \nonumber \\
	&= \frac{C'}{N} \sum_{j \in V} \sum_{t_j} c_{t_i t_j} n_{t_j} \nonumber \\
	&= \frac{C'}{N} \sum_{j \in V} c \nonumber \\
	&= C'c \label{eq: neglibible field} \, .
\end{align}
The result of \Cref{eq: neglibible field} implies that the messages in \Cref{eq:feasible_MP_1} read
\begin{align*}
q_{i \rightarrow e}(t_i)
    &\propto n_{t_i} \left(
		\prod_{\substack{f \in E \\ f \in \partial i \setminus e}} \sum_{t_j: j \in \partial f \setminus i} \pi_f \prod_{j \in \partial f \setminus i} \mess{j}{f}{t_j}
	\right) \\
	&= n_{t_i} \left(
		\prod_{\substack{f \in E \\ f \in \partial i \setminus e}} \sum_{t_j: j \in \partial f \setminus i} \pi_f \prod_{j \in \partial f \setminus i} n_{t_j}
	\right) \\
	&\propto n_{t_i} \left(
		\prod_{\substack{f \in E \\ f \in \partial i \setminus e}} \sum_{r < s \in f} \sum_{t_j: j \in \partial f \setminus i} p_{t_r t_s} \prod_{j \in \partial f \setminus i} n_{t_j}
	\right) \\
	&\propto n_{t_i} \left[
		\prod_{\substack{f \in E \\ f \in \partial i \setminus e}}
		\left(
			 \sum_{r < s \in f} \sum_{t_r t_s} p_{t_r t_s} n_{t_r} n_{t_s} + 
			 \sum_{r \in f \setminus i} \sum_{t_r} p_{t_r t_i} n_{t_r}
		\right)
	\right] \\
	&\propto n_{t_i} \left[
		\prod_{\substack{f \in E \\ f \in \partial i \setminus e}}
		\left(
			 \frac{|f|(|f|-1)}{2} c + (|f|-1)c
		\right)
	\right] \\
	&\propto n_{t_i} \, , \label{eq: external field is constant}
\end{align*}
which is exactly \cref{eq: fixed_point_2}.

\subsection{Transition matrix formula}
\label{apxsec: phase transition derivations}
In this section, we derive the expression for the transition matrix  $\tilde T_r^{ab}$ in \Cref{eq: transition matrix}. To simplify the notation, we indicate the (variable node,  function node) pairs at level $r$ as $(i_r, f_r) = (i,e)$, and similarly, at level $r+1$ we use $(i_{r+1}, f_{r+1}) = (j, f)$. Hence, the transition matrix becomes
\begin{align*}
    \tilde T_r^{ab} 
= \frac{\partial \mess{i}{e}{a}}{\partial \mess{j}{f}{b}} \,.
\end{align*}
In order to find a closed-form expression of $\tilde T_r^{ab}$, we claim that the two following Lemmas hold.
\begin{lemma}
\label{th: sum and product lemma}
Under the constant group degree assumption in \Cref{eq: constant degree assumption}:
\begin{enumerate}
\item\label{item: complex expression1} for any hyperedge $e$ and nodes $i  \in e$:
\begin{equation}\label{eq: lemma formula 1}
\sum_{t_j: j \in e \setminus i} \pi_e \prod_{k \in e \setminus i} q_{k \rightarrow e}(t_k) 
	= \frac{c|e|(|e|-1)}{2N}  \, ;
\end{equation}
\item\label{item: complex expression2} for any hyperedge $e$ and nodes $i, j \in e$:
\begin{equation}\label{eq: lemma formula 2}
\sum_{t_k: k \in e \setminus i, j} \pi_e \prod_{m \in e \setminus i, f} q_{m \rightarrow e}(t_m) 
	= \frac{1}{N} \left[
		c_{t_i t_j} 
		+ c(|e|-2) \left( 2+\frac{|e|-3}{2} \right)
	\right] \, .
\end{equation}
\end{enumerate} 
\end{lemma}

\begin{lemma}[Employing \Cref{th: sum and product lemma}]
\label{th: derivatives on h and Z}
Under the constant group degree assumption in \Cref{eq: constant degree assumption}:
\begin{enumerate}
\item\label{item: derivative h negligible} the derivative ${\partial \exp({-h (a)})} / {\partial q_{i \rightarrow e}(b)}$ is negligible to leading order in $N$;
\item\label{item: h constant} the external field is constant $h(t_i) = \const$;
\item \label{item: Z expressions} call $Z^{i \rightarrow e}$ the normalizing constant of $q_{i \rightarrow e}$, then
\begin{align}
Z^{i \rightarrow e} 
	&= \prod_{g \in \partial i \setminus e} c \frac{|g|(|g|-1)}{2N} \label{eq: normalizing constant} \\
\frac{\partial Z^{i \rightarrow e}}{\partial q_{j \rightarrow f}(b)} 
	&= \frac{c}{N} 
	\left(
		\prod_{g \in \partial i \setminus e, f} c \frac{|g|(|g|-1)}{2N}
	\right)
	\left[
		1 + (|f|-2)\left( 2 + \frac{|f|-3}{2} \right)
	\right] \, . \label{eq: normalizing constant derivative}
\end{align}
\end{enumerate}
\end{lemma}

The claims allow us to derive the transition matrix. Particularly, we make explicit all derivatives and variable-to-function nodes messages as in \Cref{eq:feasible_MP_1}. By also ignoring all terms relative to $h(t_i)$ thanks to \Cref{th: derivatives on h and Z}, we get 
\begin{align*}
\frac{\partial q_{i \rightarrow e}(a)}{\partial q_{j \rightarrow f}(b)}
	&\approx -\frac{1}{(Z^{i \rightarrow e})^2} \frac{\partial Z^{i \rightarrow e}}{\partial q_{j \rightarrow f}(b)}n_{t_i} 
	\left(
		\prod_{g \in \partial i \setminus e} \sum_{t_m : m \in \partial g \setminus i} \pi_g \prod_{m \in \partial g \setminus i} q_{m\rightarrow g}(t_m)
	\right) \\
	&+ \frac{1}{Z^{i \rightarrow e}} n_{t_i} 
	\left(
		\prod_{g \in \partial i \setminus e, f} \sum_{t_m : m \in \partial g \setminus i} \pi_g \prod_{m \in \partial g \setminus i} q_{m\rightarrow g}(t_m)
	\right)
	\left(
		\sum_{t_m : m \in \partial f \setminus i, j} \pi_f \prod_{m \in \partial f \setminus i, j} q_{m\rightarrow f}(t_m)
	\right) \, .
\end{align*}
The terms involving $Z^{i \rightarrow e}$ are in \Cref{th: derivatives on h and Z} (\Cref{eq: normalizing constant} and \Cref{eq: normalizing constant derivative}), while the expressions in parentheses are in \Cref{th: sum and product lemma} (\Cref{eq: lemma formula 1} and \Cref{eq: lemma formula 2}). By performing all the substitutions we get
\begin{align}
&\frac{\partial q_{i \rightarrow e}(a)}{\partial q_{j \rightarrow f}(b)} =\\
	&- n_{t_i}{\left(  \prod_{g \in \partial i \setminus e} c \frac{|g|(|g|-1)}{2N}  \right)^{-2}}
		\left( \prod_{g \in \partial i \setminus e, f} c \frac{|g|(|g|-1)}{2N}  \right)
	\left(
		\prod_{g \in \partial i \setminus e} c \frac{|g|(|g|-1)}{2N}
	\right)		
		\frac{c}{N}
		\left[ 1 + (|f|-2) \left( 2+ \frac{|f|-3}{2}\right)\right] \nonumber \\ 
   &+ n_{t_i}{ \left( \prod_{g \in \partial i \setminus e} c \frac{|g|(|g|-1)}{2N} \right)^{-1} }
	\left( \prod_{g \in \partial i \setminus e, f} c \frac{|g|(|g|-1)}{2N}  \right) 
	\frac{1}{N}
	\left[
		c_{t_i t_j} +c (|f|-2) \left( 2 + \frac{|f|-3}{2}\right)
	\right]  \nonumber \\
	&= {n_{t_i}}{ \left( c \frac{|f|(|f|-1)}{2N} \right)^{-1}} \left\{ -
	\frac{c}{N}
	\left[ 1 + (|f|-2) \left( 2+ \frac{|f|-3}{2}\right)\right] + 
	\frac{1}{N}
	\left[
		c_{t_i t_j} +c (|f|-2) \left( 2 + \frac{|f|-3}{2}\right)
	\right] \right\} \nonumber  \\
	&= \frac{2}{|f|(|f|-1)} n_{t_i} \left( \frac{c_{t_i t_j}}{c} -1 \right) \nonumber  \\
	&= \frac{2}{|f|(|f|-1)} n_a \left( \frac{c_{ab}}{c} -1 \right)  \nonumber  \, 
\end{align}
which is exactly the expression in \cref{eq: transition matrix}.

What is left to complete all derivations is to prove \Cref{th: sum and product lemma} and \Cref{th: derivatives on h and Z}, which is done next.

\subsubsection{Proof of \Cref{th: sum and product lemma}}

\begin{itemize}
    \item[1.] Derivation of \Cref{eq: lemma formula 1}:
    \begin{align*}
    \sum_{t_j: j \in e \setminus i} \pi_e \prod_{k \in e \setminus i} q_{k \rightarrow e}(t_k) 
    	&= \sum_{t_j: j \in e \setminus i} \sum_{r<s \in e} p_{t_r t_s} \prod_{k \in e \setminus i} q_{k \rightarrow e}(t_k) \\
    	&= \sum_{r<s \in e\setminus i} \sum_{t_r t_s} p_{t_r t_s} q_{r \rightarrow e}(t_r) q_{s \rightarrow e}(t_s) + \sum_{r \in e\setminus i} \sum_{t_r} p_{t_r t_i} q_{r \rightarrow e}(t_r) \\
    	&= \frac{1}{N}\sum_{r<s \in e\setminus i} \sum_{t_r, t_s} c_{t_r t_s} n_{t_r} n_{t_s}  + \frac{1}{N}\sum_{r \in e\setminus i} \sum_{t_r} c_{t_r t_i} n_{t_r} \\
    	&= \frac{c}{N}\left[
    		\frac{(|e|-1)(|e|-2)}{2} + (|e|-1 )
    	\right] \\
    	&= \frac{c(|e|-1)|e|}{2N} \, .
    \end{align*}
    \item[2.] Derivation of \Cref{eq: lemma formula 2}:
    \begin{align}
    \fl
    \sum_{t_k: k \in e \setminus i, j} \pi_e \prod_{m \in e \setminus i, j} q_{m \rightarrow e}(t_m) 
    	&= \sum_{t_k: k \in e \setminus i, j} \sum_{r<s \in e} p_{t_r t_s} \prod_{m \in e \setminus i, j} q_{m \rightarrow e}(t_m) \nonumber  \\
    	&= p_{t_i t_j}+ \sum_{ r \in e\setminus i, j}\sum_{t_r} p_{t_r t_i} q_{r \rightarrow e}(t_r) + \sum_{ r \in e\setminus i, j}\sum_{t_r} p_{t_r t_j} q_{r \rightarrow e}(t_r) \nonumber \\
        & + \sum_{r<s \in e \setminus i, j} \sum_{t_r, t_s} p_{t_r t_s} q_{r \rightarrow e}(t_r)q_{s \rightarrow e}(t_s) \nonumber \\
    	&= \frac{1}{N} \Bigg( c_{t_i t_j} + \sum_{r  \in e\setminus i, j}\sum_{t_r} c_{t_r t_i} n_{t_r} + \sum_{ r \in e\setminus i, j}\sum_{t_r} c_{t_r t_j} n_{t_r} + \sum_{r<s \in e \setminus i, j}\sum_{t_r, t_s} c_{t_r t_s} n_{t_r} n_{t_s} \Bigg) \nonumber \\
    	&= \frac{1}{N} \left( c_{t_i t_j} + \sum_{ r \in e\setminus i, j} c + \sum_{ r \in e\setminus i, j}c + \sum_{r<s \in e \setminus i, j}c \right) \nonumber  \\
    	&= \frac{1}{N} \left[
    		c_{t_i t_j} 
    		+ c(|e|-2) \left( 2+\frac{|e|-3}{2} \right)
    	\right] \, \nonumber .
    \end{align}
\end{itemize}

\subsubsection{Proof of \Cref{th: derivatives on h and Z}}

\begin{itemize}
\item[1.] Using \Cref{eq:feasible_MP_4}, we write
\begin{align}
\label{eq: non-zero derivatives}
\frac{\partial \exp({-h (a)})}{\partial q_{i \rightarrow e}(b)} 
	= \exp\left(- \frac{C'}{N} \sum_{v \in V} \sum_{t_k} c_{a t_k} q_k(t_k)\right)
	\left(
	- \frac{C'}{N} \sum_{k \in V} \sum_{t_v} c_{a t_k} \frac{\partial q_k(t_k)}{\partial q_{i \rightarrow e}(b)}
	\right) \, .
\end{align}
Only few of the derivatives ${\partial q_k(t_k)}/{\partial q_{i \rightarrow e}(b)}$ entering \Cref{eq: non-zero derivatives} are non-zero. Hence, the full derivative has negligible order $O(1/N)$.
\item[2.] The fact that the external field is constant was already shown in \Cref{eq: neglibible field} during the proof of \Cref{th: constant degree and message fixed points}.
\item[3.] As just proved, we can ignore the external field in the expression of $Z^{i \rightarrow e}$, and find
\begin{align}
Z^{i \rightarrow e} 
	&\approx \sum_{t_i} q_{i \rightarrow e}(t_i) \nonumber \\
    \label{eq: to simplify in lemma}
	&= \sum_{t_i} n_{t_i} \left(\prod_{g \in \partial i \setminus e} \sum_{t_j: j \in \partial g \setminus i} \pi_g \prod_{j \in \partial g \setminus i} q_{j \rightarrow g}(t_j) \right) \, .
\end{align}
Utilizing result \Cref{eq: lemma formula 1} in \Cref{th: sum and product lemma}, \Cref{eq: to simplify in lemma} simplifies to 
\begin{align*}
Z^{i \rightarrow e} 
	&= \sum_{t_i} n_{t_i} \left(\prod_{g \in \partial i \setminus e} c\frac{|g|(|g|-1)}{2N} \right) \\
	&=  \left(\prod_{g \in \partial i \setminus e} c\frac{|g|(|g|-1)}{2N} \right) \, .
\end{align*}
which results in \Cref{eq: normalizing constant}, as desired. Similarly, to compute the derivative ${\partial Z^{i \rightarrow e}} / {\partial q_{j \rightarrow f}(b)} $ we can ignore all appearing ${\partial \exp({-h (a)})} / {\partial q_{j \rightarrow f}(b)}$ and $h(t_i)$ thanks the Lemma's first two points (just proved). Hence
\begin{align*}
\frac{\partial Z^{i \rightarrow e}}{\partial q_{j \rightarrow f}(b)} 
	&= \frac{\partial}{ \partial q_{j \rightarrow f}(b)} 
		 \left[ \displaystyle  \sum_{t_i} n_{t_i} \left(\prod_{g \in \partial i \setminus e} \sum_{t_j: j \in \partial g \setminus i} \pi_g \prod_{j \in \partial g \setminus i} q_{j \rightarrow g}(t_j) \right) \right]\\
	&=\sum_{t_i} n_{t_i} 
	\left(
		\prod_{g \in \partial i \setminus e, f} \sum_{t_m: m \in \partial g \setminus i} \pi_g \prod_{m \in \partial g \setminus i} q_{m \rightarrow g}(t_m)
	\right)
	\left(
		\sum_{t_m: m \in \partial f \setminus i, j} \pi_f \prod_{m \in \partial f \setminus i, j} q_{m \rightarrow f}(t_m)
	\right) \, ,
\end{align*}
and using \Cref{eq: lemma formula 1} and \Cref{eq: lemma formula 2} from \Cref{th: sum and product lemma}, conclude with
\begin{align*}
&=  \frac{1}{N}\left(
		\prod_{g \in \partial i \setminus e, f} c \frac{|g|(|g|-1)}{2N}
	\right)
	\left[
		 \sum_{t_i} n_{t_i} c_{t_i t_j} + c(|f|-2)\left(2 + \frac{|f|-3}{2}	\right) 
	\right] \\
&=  \frac{c}{N}\left(
		\prod_{g \in \partial i \setminus e, f} c \frac{|g|(|g|-1)}{2N}
	\right)
	\left[
		 1+ (|f|-2)\left(2 + \frac{|f|-3}{2}	\right) 
	\right] \, .
\end{align*}
\end{itemize}

\subsection{Elapsed time of MP}
\label{apxsec: Elapsed time of MP}

\begin{figure}[htpb]
    \centering
    \includegraphics[width=0.5\columnwidth]{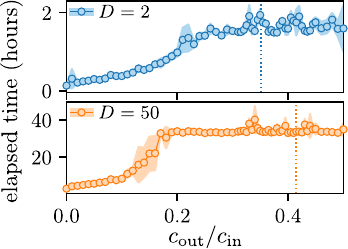}
    \caption{
    \textbf{Elapsed time for MP.} For both $D=2$ and $D=5$, the elapsed times plateau due to the threshold imposed on MP's maximum number of iterations. Shaded areas are standard deviations over $5$ random initializations of MP. Vertical dotted lines are theoretical detectability bounds derived from \Cref{eq: stability criterion}.
    }
    \label{fig:elapsed_time}
\end{figure}

In \Cref{fig:elapsed_time}, we plot the running time of MP when performing the synthetic experiments of \Cref{subsec: phase transition in hyg}. Elapsed times become prohibitively large when $c_{\mathrm{out}}/c_{\mathrm{in}}$ increases. For this reason, we threshold the maximum number of MP iterations and obtain the plateaus of \Cref{fig:elapsed_time}.

\section{Calculations of the free energy}
\label{apxsec: free energy}

After MP, it is possible to approximate the log-evidence of the data, i.e., the log-normalizing constant $\log Z$, as per \Cref{eq: gibbs_distribution}. The equivalent quantity $F = - \log Z$, called the free energy of the system, can be obtained via the following cavity-based general formula:
\begin{equation}\label{eq: free energy cavity approx}
    F \approx -\sum_{i \in V} f_i + \sum_{e \in \Omega} (|e|-1) f_e \, ,
\end{equation}
where 
\begin{align*}
    f_i &= \log\left(
        \sum_{t_i} n_{t_i} \prod_{e \in \partial i} \sum_{t_j: j \in \partial e \setminus i} \left( \frac{\pi_e}{\kappa_e} \right)^{A_e} \left( 1- \frac{\pi_e}{\kappa_e} \right)^{1-A_e} \prod_{j \in \partial e \setminus i} \mess{j}{e}{t_j} 
    \right) \\
    f_e &= \log\left(
        \sum_{t_j: j \in \partial e} \left( \frac{\pi_e}{\kappa_e} \right)^{A_e} \left( 1- \frac{\pi_e}{\kappa_e} \right)^{1-A_e} \prod_{j \in \partial e} \mess{j}{e}{t_j} 
    \right) \, .
\end{align*}
Assuming that MP has converged, all messages $\mess{j}{e}{t_j}$ are available. Notice, however, that naive computations of the $f_i$ and $f_e$ addends are unfeasible, due to the exploding sums over ${t_j: j \in \partial e}$. In the following, we show how such computations can be performed efficiently. 

\begin{enumerate}
    \item Calculations of $f_i$.
    
    As one can observe from \Cref{eq: messtwo formula appendix} and \Cref{eq: general marginals appendix}, the $f_i$ terms are the log-normalizing constants of $q_i$, therefore they can be computed similarly. In particular, ignoring constants, by \Cref{eq: ext field}, the following simplification holds:
\begin{equation*}
    f_i = \log \left(
        \sum_{t_i} n_{t_i} \prod_{\substack{e \in E \\ e \in \partial i}} \sum_{t_j: j \in \partial e \setminus i} \pi_e \prod_{j \in \partial e \setminus i} \mess{j}{e}{t_j} 
        \exp( - h(t_i))
    \right)
    \, .
\end{equation*}
The single terms indexed by $e \in E$, i.e., the values 
$\sum_{t_j: j \in \partial e \setminus i} \pi_e \prod_{j \in \partial e \setminus i} \mess{j}{e}{t_j} $, 
are equivalent to the unnormalized messages $\messtwo{e}{j}{t_j}$. For this reason, they can be computed with the same dynamic program presented in \ref{apxsec: dynamic programming}.

\item Calculations of $f_e$.

{While the $f_i$ terms in \Cref{eq: free energy cavity approx} are computed singularly, we take a different approach and calculate the whole sum $\sum_{e \in \Omega} (|e|-1) f_e$ without computing the single $f_e$, as this would be impossible due to their exploding number.
First, we separate the terms over $\Omega$ in \Cref{eq: free energy cavity approx} as follows.
\begin{align*}
\sum_{e \in \Omega} (|e|-1) f_e
    &= \sum_{e \in \Omega} (|e|-1) \log\left(
        \sum_{t_j: j \in \partial e} \left( \frac{\pi_e}{\kappa_e} \right)^{A_e} \left( 1- \frac{\pi_e}{\kappa_e} \right)^{1-A_e} \prod_{j \in \partial e} \mess{j}{e}{t_j} 
    \right) \\
    &= \log \left(
        \prod_{e \in \Omega} \left[
            \sum_{t_j: j \in \partial e} \left( \frac{\pi_e}{\kappa_e} \right)^{A_e} \left( 1- \frac{\pi_e}{\kappa_e} \right)^{1-A_e} \prod_{j \in \partial e} \mess{j}{e}{t_j} 
        \right]^{|e|-1}
    \right) \\
    &= \log \left(
        \prod_{e \in E} \left[
            \sum_{t_j: j \in \partial e} \pi_e \prod_{j \in \partial e} \mess{j}{e}{t_j} 
        \right]^{|e|-1}
    \right) + 
    \log \left(
        \prod_{e \in \Omega \setminus E} \left[
            \sum_{t_j: j \in \partial e} \left( 1- \frac{\pi_e}{\kappa_e} \right) \prod_{j \in \partial e} \mess{j}{e}{t_j} 
        \right]^{|e|-1}
    \right) +  \const \, .
\end{align*}
This allows us to compute the last two addends separately.

Focusing on the second addend, and proceeding similarly as for the external field calculations that brought to \Cref{eq: external field final apx}, we get
\begin{align*}
   \log \prod_{e \in \Omega \setminus E} &\left[
        \sum_{t_j: j \in \partial e} \left( 1- \frac{\pi_e}{\kappa_e} \right) \prod_{j \in \partial e} \mess{j}{e}{t_j} 
    \right]^{|e|-1}\\
    &\approx \log  \prod_{e \in \Omega \setminus E} \left[ \exp \left(
        -\frac{1}{N} \sum_{t_j: j \in \partial e} \left( \frac{\sum_{k < m \in e} c_{t_k t_m}}{\kappa_e} \right) \prod_{j \in \partial e} \mess{j}{e}{t_j} 
    \right) \right]^{|e|-1} \\
    &\approx \log  \prod_{e \in \Omega \setminus E} \left[ \exp \left(
        -\frac{1}{N} \sum_{t_j: j \in \partial e} \left( \frac{\sum_{k < m \in e} c_{t_k t_m}}{\kappa_e} \right) \prod_{j \in \partial e} q_j(t_j)
    \right) \right]^{|e|-1} \\
    &= \log \prod_{e \in \Omega \setminus E} \exp \left(
        \frac{1-|e|}{N} \sum_{t_j: j \in \partial e} \left( \frac{\sum_{k < m \in e} c_{t_k t_m}}{\kappa_e} \right) \prod_{j \in \partial e} q_j(t_j)
    \right) \\
    &\approx \log \prod_{e \in \Omega} \exp \left(
        \frac{1-|e|}{N} \sum_{t_j: j \in \partial e} \left(\frac{\sum_{k < m \in e} c_{t_k t_m}}{\kappa_e} \right) \prod_{j \in \partial e} q_j(t_j)
    \right) \\
    &= \sum_{e \in \Omega} 
        \frac{1-|e|}{N} \sum_{t_j: j \in \partial e} \left(\frac{\sum_{k < m \in e} c_{t_k t_m}}{\kappa_e} \right) \prod_{j \in \partial e} q_j(t_j) 
    \\
    &= \frac{1}{N}\sum_{e \in \Omega} 
        \frac{1-|e|}{\kappa_e} \sum_{k < m \in e} \sum_{t_j: j \in \partial e} c_{t_k t_m} \prod_{j \in \partial e} q_j(t_j)
    \\
    &= \frac{1}{N}\sum_{e \in \Omega} 
        \frac{1-|e|}{\kappa_e} \sum_{k < m \in e} \sum_{t_k t_m} c_{t_k t_m} q_k(t_k) q_j(t_j)
    \\
    &= \frac{C'''}{N} \sum_{k < m \in V} \sum_{t_k t_m} c_{t_k t_m} q_k(t_k) q_j(t_j) \, 
\end{align*}
where $C''' := \sum_{d=2}^D \frac{1-d}{\kappa_d} \binom{N-2}{d-2}$. 
Also define $q_V(a) = \sum_{k \in V} q_k(a)$. Then,
\begin{align*}
\log \prod_{e \in \Omega \setminus E} \left[
        \sum_{t_j: j \in \partial e} \left( 1- \frac{\pi_e}{\kappa_e} \right) \prod_{j \in \partial e} \mess{j}{e}{t_j} 
    \right]^{|e|-1}
    &\approx \frac{C'''}{N} \, \sum_{k < m \in V} \sum_{t_k t_m} c_{t_k t_m} q_k(t_k) q_j(t_j) \\
    &= \frac{C'''}{N} \, \sum_{a b} c_{ab} \sum_{k < m \in V} q_k(a) q_j(b) \\
    &= \frac{C'''}{2 \, N} \, \sum_{a b} c_{ab} \left[
        \sum_{k, m \in V} q_k(a) q_j(b) - \sum_{k \in V} q_k(a) q_k(b)
    \right] \\
    &= \frac{C'''}{2 \, N} \, \sum_{a b} c_{ab} \left[
        q_V(a) q_V(b) - \sum_{k \in V} q_k(a) q_k(b)
    \right] \,
\end{align*}
which can be computed in linear time $O(|V| \, K^2)$.

The first addend requires different considerations. Since naive calculations of every sum on ${t_j: j \in \partial e}$ cost $O(K^{|e|})$, and thus are unfeasible, we design a dynamic program similar to that of \ref{apxsec: dynamic programming}. For simplicity, consider a hyperedge $e={1, \ldots, m}$. Proceeding as in \ref{apxsec: dynamic programming}, we define the quantities:
\begin{align}
    \label{eq: dyn prog lik 1}
    \tilde{\eta}(e, n) &= \sum_{t_j: j =1, \ldots, n} \pi_e \prod_{j =1, \ldots, n} \mess{j}{e}{t_j} \\
    \label{eq: dyn prog lik 2}
    \tilde{s}_n(a) &= \sum_{t_1} p_{a t_1} \mess{1}{e}{t_1} + \ldots + \sum_{t_{n-1}} p_{a t_{n-1}} \mess{n-1}{e}{t_{n-1}} \, .
\end{align}
Notice that $\tilde{\eta}(e, m) = \sum_{t_j: j \in \partial e} \pi_e \prod_{j: j \in \partial e} \mess{j}{e}{t_j}$ is the quantity we need to compute. For \Crefrange{eq: dyn prog lik 1}{eq: dyn prog lik 2}, the following recursions hold:
\begin{align*}
    \tilde{\eta}(e, n) &= \tilde{\eta}(e, n-1) + \sum_{t_n} \mess{n}{e}{t_n} \tilde{s}_n(t_n) \\
    \tilde{s}_n(a) &= \tilde{s}_{n-1}(a) + \sum_{t_{n-1}} p_{t_{n-1}a} \mess{n-1}{e}{t_{n-1}} \, .
\end{align*}
Here, computing the final $\tilde{\eta}(e, m)$ costs $O(K |e|)$, and computing it for all the observed hyperedges costs $\sum_{e \in E} O(K |e|)$. Notice that utilizing the dynamic program in \ref{apxsec: dynamic programming} would cost $\sum_{e \in E} O(K^2 |e|^2)$, plus the processing needed to obtain the $\tilde{\eta}(e, m)$ value. Hence, the new recursions result in good computing savings with minimal changes to the numerical implementation.}
\end{enumerate}

\subsection{Computation of the free energy landscape on High School data}
\label{apxsec: free energy on high school}

We explain further how to obtain the free energy landscape of the High School dataset in \Cref{fig: real exps}. 
The three vertices are inferred using the dataset's hyperedges whose size is lower or equal than $D$, with $D=2, 3, 4$.  After having performed inference on every vertex, we obtain the parameters $(p_2, n_2), (p_3, n_3), (p_4, n_4)$---each pair is associated with a value of $D$---for the affinity matrix and the community prior.

Every point in the simplex is generated with a convex combination of the three vertices.  Particularly, we define the parameters
\begin{align*}
	p_{\mathrm{simplex}} &= \lambda_2 p_2 + \lambda_3 p_3 + \lambda_4 p_4 \\
	n_{\mathrm{simplex}} &= \lambda_2 n_2 + \lambda_3 n_3 + \lambda_4 n_4 \, ,
\end{align*}
where $0 \leq \lambda_i \leq 1$ and $\sum_{i={2,3,4}} \lambda_i = 1$. For any value of $p_{\mathrm{simplex}},  n_{\mathrm{simplex}}$, we compute the free energy on the whole High School dataset, i.e., taking all hyperedges. The free energy approximations following \Cref{eq: free energy cavity approx} require the messages, marginals and external field, which can be inferred via MP and in turn depend on $p_{\mathrm{simplex}},  n_{\mathrm{simplex}}$. For every point in the simplex, we fix $p_{\mathrm{simplex}},  n_{\mathrm{simplex}}$ and infer all the remaining quantities via MP, to then compute the free energy displayed in \Cref{fig: real exps}.

\subsection{Inference of class affinity on High School data}
\label{apxsec: affinity on high school}

\begin{figure}[htpb]
\centering
\includegraphics[width=1\columnwidth]{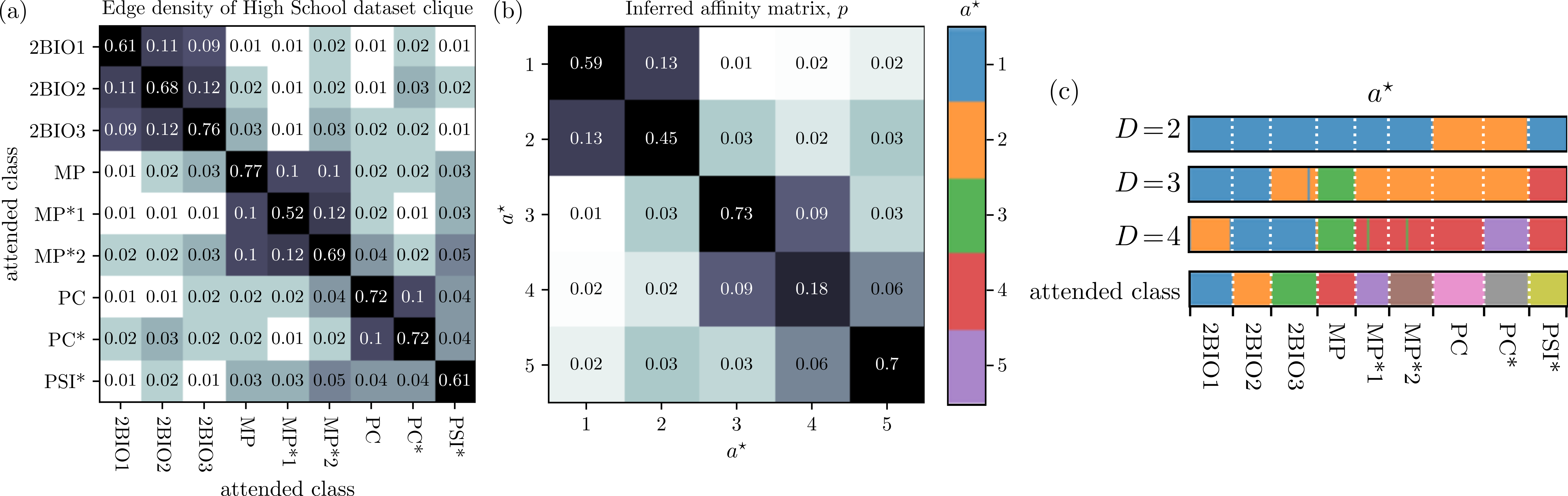}
\caption{\textbf{Affinity patterns on the High School dataset.} Colors of the matrices' entries correspond to their log values, properly normalized to ease the Figure's readability. \textbf{(a)} Edge density on the clique decomposition of the High School dataset. As in Mastandrea \emph{et al.} \cite{mastrandrea2015contact}, the edge density between two classes $X$ and $Y$ corresponds to the number of observed edges between nodes of the classes, normalized with respect to the total number of possible edges between $X$ and $Y$.  \textbf{(b)} Affinity matrix $p$ inferred by the EM-MP scheme with $D=4$. The method detects $5$ classes, whose affinity values are as in the matrix's entries. Colors of classes follow the color coding of \Cref{fig: real exps}(c). \textbf{(c)} Inferred communities of nodes and the partition in the classes of students. The panel is identical to \Cref{fig: real exps}(c).}
\label{fig: inference affinity high school}
\end{figure}

We expand on the community patterns detected on the High School data for $D=4$, which are represented in \Cref{fig: real exps}. The nine classes observed in the data are named after their subjects of focus, and are: MP,  MP*1,  MP*2 (mathematics and physics), PC, PC* (physics and chemistry), PSI* (engineering), 2BIO1, 2BIO2, 2BIO3 (biology) \cite{mastrandrea2015contact}.

We compare the the edge density patterns computed on the data in Mastandrea \emph{et al.} \cite{mastrandrea2015contact}, and shown in \Cref{fig: inference affinity high school}(a), with the affinity matrix $p$ inferred on the High School dataset fixing $D=4$, shown in \Cref{fig: inference affinity high school}(b).  Additionally,  in \Cref{fig: inference affinity high school}(c), we plot the partition of the nodes into communities with their labeling in classes.

We observe that classes that are inferred in the same community appear to also belong to classes that have a larger number of external interactions with other classes in the same inferred community. For instance, the BIO classes belong to two communities that are disjoint from all others, see \Cref{fig: inference affinity high school}(c). Within the BIO classes,  2BIO2 and 2BIO3 are grouped in the same community,  as they have slightly higher edge density of $0.12$, compared to the $0.11$ and $0.09$ observed for 2BIO1. 

The affinity matrix shown in \Cref{fig: inference affinity high school}(b) aligns well with the inter- and intra-community interactions. For instance, communities $1$ and $2$ (that contain the BIO classes) have an upper diagonal block that isolates them from all others.  Communities $3$ and $5$, which largely match students from classes MP and PC, are disassortative with the remaining classes, grouped in community $4$. 

\subsection{Further comments on higher-order interactions on High School data}
\label{apxsec: affinity on high school 2}
\begin{table}[hptb]
\small
    \centering
    \setlength{\tabcolsep}{3pt}
    \setlength{\arrayrulewidth}{0.1pt}
    \begin{tabular}{cc}
\toprule
$D$ & AUC \\
\midrule 
$2$ & $0.710 \pm 0.002$ \\
$3$ & $0.780 \pm 0.003$ \\
$4$ & $0.843 \pm 0.004$ \\
$5$ & $0.813 \pm 0.003$ \\
\bottomrule
\end{tabular}
    \caption{\textbf{AUC scores on the High School dataset}.  We perform MP and EM inference on the High School dataset utilizing hyperedges up to size $D$. Then, we compute the AUC on the full dataset, i.e. , on the hypergraph with all hyperedges up to $D=5$. The goodness of link prediction, represented by the AUC score, shows that interactions up to size $4$ improve the quality of inference, while utilizing interaction of size $5$ yields a slight drop in performance.}
    \label{tab:AUC high school}
\end{table}

The High School hypergraph contains interactions of orders ranging from $2$ to $5$. In our experiments, we observe that optimal inference is reached at maximum hyperedge size $D=4$, while utilizing interactions of order $5$ slightly degrades performance.  We confirm this in varioys ways.  The communities inferred (now shown) are less granular than those presented for $D=4$. A similar trend is observed in the free energy (not shown), that slightly increases when performing inference on the whole dataset.
Finally, we measure the link prediction performances utilizing parameters inferred with $D=2, 3, 4, 5$, and compute the AUC with respect to the full dataset, which we include in \Cref{tab:AUC high school}.  Here again we observe a slight drop in AUC when utilizing parameters inferred at $D=5$, despite the AUC being computed with respect to all hyperedges, including those not observed when training on lower values of $D$.

There could be various reasons for this result. A possible explanation is that the interactions at $D=5$ are noisier and/or less aligned with the data generating process assumed by our generative model. We recall that the data is collected via proximity sensors, and that social interactions in larger groups are harder to detect, and may arise from different types of link formation mechanisms. 

\bibliographystyle{iopart-num}
\bibliography{bibliography}

\begin{thebibliography}{10}
\bibitem{gauss_corr}
To deal with large $N_{\#}$ and $\kappa_d$ that cannot be stored in memory, we
  approximate the binomial in \Cref{eq: binomial count of samples hyperdges}
  with a Gaussian \begin{equation*} \mathcal{N} \left(
  \frac{\pi_{\#}}{\kappa_d} N_{\#}, \frac{\pi_{\#}}{\kappa_d} N_{\#} \left( 1-
  \frac{\pi_{\#}}{\kappa_d} \right) \right) \, \end{equation*} Crucially, the
  Gaussian's mean and variance only involve the ratio $N_{\#} / \kappa_d$,
  which is numerically stable. We adopt this approximation when the Gaussian's
  variance exceeds $10$.
\bibitem{pois_corr}
A Poisson approximation of the binomial \begin{align*} \mathrm{Pois}\left(
  N_{\#} \frac{\pi_{\#}}{\kappa_d} \right) \end{align*} is used if $N_{\#} >
  20$ and ${N_{\#} \pi_{\#}} / {\kappa_d} < 0.1$, or if $N_{\#} > 100$ and
  ${N_{\#} \pi_{\#}} / {\kappa_d} < 10$.
\end{thebibliography}

\end{document}